\documentclass[letterpaper,11pt]{article}

 \hoffset= - 0.75 in

\title{\bf Toric Ideals of Phylogenetic Invariants}

\author{ {\bf Bernd Sturmfels \  and \ Seth Sullivant}
 \\ {\small Department of Mathematics, University of California,
Berkeley} }

\date{}

\usepackage{latexsym,array,delarray,amsthm,amssymb}
\usepackage[sumlimits]{amsmath}
\usepackage[dvips]{graphicx}

\theoremstyle{plain}
\newtheorem{thm}{Theorem}
\newtheorem{lemma}[thm]{Lemma}
\newtheorem{prop}[thm]{Proposition}

\newtheorem{conj}[thm]{Conjecture}

\theoremstyle{definition}
\newtheorem{defn}[thm]{Definition}
\newtheorem{ex}[thm]{Example}

\newtheorem{rmk}[thm]{Remark}

\theoremstyle{remark}

\newcommand{\pp}{\mathbb{P}}

\newcommand{\rr}{\mathbb{R}}
\newcommand{\cc}{\mathbb{C}}

\begin{document}

\maketitle
\begin{abstract}
Statistical models of evolution are algebraic varieties in the
space of joint probability distributions on the leaf colorations
of a phylogenetic tree. The phylogenetic invariants of a model are
the polynomials which vanish on the variety. Several widely used
models for biological sequences have transition matrices that can
be diagonalized by means of the Fourier transform of an abelian
group. Their phylogenetic invariants form a toric ideal in the
Fourier coordinates. We determine generators and Gr\"obner bases
for these toric ideals. For the Jukes-Cantor and Kimura models on
a binary tree, our Gr\"obner bases consist of certain explicitly
constructed  polynomials of degree at most four.
\end{abstract}

\section{Introduction}

Cavender-Felsenstein \cite{CF} and Lake \cite{La} introduced
phylogenetic invariants as an algebraic tool for reconstructing
evolutionary trees from biological sequence data. Such invariants
exist for any tree-based Markov model, and they uniquely
characterize that model. While partial lists of invariants have
been described for various models \cite{EZ, FS1, Ha, SB}, the
literature still conveys a sense that phylogenetic invariants and
algebraic algorithms for computing them  are not useful for any
problem whose size is of biological interest. In his book {\sl
Inferring Phylogenies}, Felsenstein sums this up from the
perspective of molecular biology as follows: {\sl ... invariants
are worth attention, not for what they do for us now, but what
they might lead to in the future...}
 \cite[page 390]{Fel}. A similar tone is expressed
in the final section of the book {\sl Phylogenetics} by Semple and
Steel \cite[page 212]{SS}.

But the future is closer than readers of these two excellent books
might surmise. For the general Markov model, considerable progress
has been made in the recent work of Allman and Rhodes \cite{AR1,
AR2}. A determinantal representation of the Allman-Rhodes ideal in
the binary case has been proposed in
 \cite[\S 5]{PS}.
The present paper is not concerned with the general Markov model
but with a class of special models, namely, the
  \emph{group-based models}
\cite[\S 8.10]{SS}.  The problem of finding invariants for these
models was studied by several authors including
 Evans-Speed \cite{ES}, Evans-Zhou \cite{EZ},
Steel-Fu \cite{SF} and Sz\'ekely-Steel-Erd\"os \cite{SSE}. The
class of group-based models includes the \emph{Jukes-Cantor
model}, for either binary or DNA sequences, and the \emph{Kimura
models}, with  two or three parameters.

The contribution of this paper is an explicit description of
generators and a  Gr\"obner basis for the ideal of phylogenetic
invariants of such a model. The key idea can be summarized as
follows.

\begin{thm} \label{main}
For any group based model on a phylogenetic tree $T$, the prime
ideal of phylogenetic invariants is generated by the invariants of
the local submodels around each interior node of $T$, together
with the quadrics which encode conditional independence statements
along the splits of $T$.
\end{thm}

The precise form of this theorem and its proof are given in
Section 4. We continue by reviewing the evolutionary models to be
considered here. Let $T$ be a rooted tree with $m$ leaves. Let
$\mathcal{V}(T)$ denote the set of nodes of $T$.  To each node $v
\in \mathcal{V}(T)$ we associate a $k$-ary random variable $X_v$.
In biology, the most common values of $k$ are $2, 4,$ and $20$.
Consider the probability $P(X_v = i)$ that $X_v$ is in state $i$.
For DNA sequences this probability represents the proportion of
characters in the sequence at $v$ which is a particular
nucleotide, namely, $A$, $G$,  $C$ or $T$.

The relationship between the random variables $X_v$ is encoded by
the structure of the tree. Let $\pi$ be a distribution of the
random variable $X_r$ at the root node $r$.  For each node $v \in
\mathcal{V}(T) \backslash \{r\}$, let $a(v)$ be the unique parent
of $v$. The transition from $a(v)$ to $v$ is given by a $k \times
k$-matrix $A^{(v)}$ of probabilities.  Then the probability
distribution at each node is computed recursively by the rule
\begin{equation}
\label{recursive P} P(X_v = j) \quad = \quad \sum_{i=1}^k
A^{(v)}_{ij} \cdot P(X_{a(v)}=i).
\end{equation}
This rule induces a joint distribution on all the random variables
$X_v$. We label the leaves of $T$ by
 $1,2,\ldots,m $, and we abbreviate
the marginal distribution on the variables at the leaves as
follows:
\begin{equation}
\label{unknownP}
 p_{i_1 i_2 \ldots i_m} \quad = \quad
P(X_{1} = i_1,X_{2} = i_2, \ldots, X_{m} = i_m ).
\end{equation}
In biological applications, one estimates (some of) these $k^m$
probabilities from $m$ aligned sequences on $k$ letters, and the
aim is to reconstruct the tree. The root distribution $\pi$ and
the transition matrices $A^{(v)}$ are typically unknown. In the
general Markov model of \cite{AR1}, each matrix entry
$A^{(v)}_{ij}$ is an independent model parameter. For the
group-based models, to be studied in this paper, the number of
model parameters is smaller because some of the entries of
$A^{(v)}$ are assumed to coincide.

A \emph{phylogenetic invariant} of the model is a polynomial in
the leaf probabilities $p_{i_1 i_2 \cdots i_m}$ which vanishes for
every choice of model parameters. The set of these polynomials
forms a prime ideal in the polynomial ring over the unknowns
$p_{i_1 i_2 \cdots i_m}$.  Our objective is to compute this ideal
as explicitly as possible. In the language of algebraic geometry,
we seek to determine the variety parameterized by the rational map
induced by the joint distribution on the leaves. The  study of
such varieties for various statistical models is a central theme
in the emerging field of \emph{algebraic statistics}  \cite{GSS,
HS}.

In this paper, we determine the ideal of invariants for models
whose  structure is governed by an abelian group. Four models used
in computational biology have this structure: the Jukes-Cantor
models and the Kimura models. Theorem 2 below summarizes our
results for these models.

The \emph{Jukes-Cantor binary model} on two letters ($k=2$) is the
model with transition matrices
$$A^{(v)} \quad =  \quad \left( \begin{array}{cc} 1 - a_v & a_v \\
a_v & 1-a_v \end{array} \right), $$ where $a_v$ is the probability
of making a transition between the states along the edge from
$a(v)$ to $v$. The \emph{Kimura 3 parameter model} on $k=4$
letters (for DNA sequences) has the transition matrices
$$A^{(v)} \quad =  \quad \left( \begin{array}{cccc}
1 - a_v - b_v- c_v  & a_v & b_v & c_v \\
a_v & 1-a_v - b_v - c_v & c_v & b_v \\
b_v & c_v & 1-a_v- b_v- c_v & a_v \\
c_v & b_v & a_v & 1-a_v- b_v- c_v \end{array} \right) $$ \noindent
where $a_v$ is the probability of a transition and $b_v$ and $c_v$
are the transversion probabilities.  The Kimura 2-parameter model
arises as the subvariety defined by taking $b_v = c_v$ for all $v$
and the Jukes-Cantor DNA model is the subvariety defined by
setting $a_v = b_v = c_v$ for all $v$.

Evans and Speed \cite{ES} introduced a linear change of
coordinates which diagonalizes  these models.  In Section 2 we
review their Fourier transform at the level of generality proposed
by Sz\'ekely-Steel-Erd\"os \cite{SSE}.  The key idea is to label
the states of the random variables $X_v$ by a finite abelian group
(e.g. $\mathbb{Z}_2 = \{0,1\} $ or $\mathbb{Z}_2 \times
\mathbb{Z}_2 = \{A,G,C,T\}$) in such a way that
 the probability of transitioning from  $g_i$ to $g_j$ depends
only on the difference $g_i - g_j$. Replacing the original
coordinates $p_{i_1 \cdots i_m}$ by Fourier coordinates $q_{i_1
\cdots i_m}$, the ideal of phylogenetic invariants becomes a toric
ideal. Recall (e.g.~from \cite{berndbook}) that a \emph{toric
ideal} is a prime ideal generated by differences of monomials.

As an example consider the Jukes-Cantor binary model for $m=4$.
The  Fourier coordinates are
\begin{equation}
\label{fouriercoords}
 q_{ijkl} \quad = \quad
\sum_{r=0}^1 \sum_{s=0}^1 \sum_{t=0}^1 \sum_{u=0}^1
(-1)^{ir+js+kt+lu} \cdot p_{rstu} , \qquad \hbox{where} \,\,\,
i,j,k,l \,\in\, \mathbb{Z}_2.
\end{equation}
If $T$ is the balanced binary tree of height two, then this model
has the parametric representation
\begin{equation}
\label{smallmap} q_{ijkl} \quad \mapsto \quad a_i \cdot b_{j}
\cdot c_k \cdot d_{l} \cdot e_{i+j}  \cdot f_{k+l} \cdot
g_{i+j+k+l}.
\end{equation}
Disregarding the trivial invariant $\,q_{0000} -1 $, the toric
ideal of phylogenetic invariants is generated by
 $20$ linearly independent quadrics. These arise as the
$2 \times 2$-minors of the four $2 \times 4$-matrices
\begin{equation}
\label{twomatrices}
\begin{pmatrix} \!
q_{0\, i \, 00} & q_{0 \, i \, 01} & q_{0 \, i \, 10} &q_{0 \, i \, 11} \\
 q_{1 (1+i) 00} \!\! & \!\! q_{1 (1+i) 01} \!\! & \!\! q_{1 (1+i) 10}
\!\! & \!\! q_{1 (1+i) 11} \! \\
\end{pmatrix}
 \hbox{and}
\begin{pmatrix}
q_{00 \, i \, 0} & q_{01 \, i \, 0} & q_{10 \, i \, 0} & q_{11 \, i \, 0} \\
q_{00 (1+i) 1} \!\! & \! \! q_{01 (1+i) 1} \!\! & \!
\! q_{10 (1+i) 1} \!\! & \!\! q_{11 (1+i) 1} \! \\
\end{pmatrix}
\, \hbox{for}\, i = 0,1.
\end{equation}
Moreover, these quadrics form a Gr\"obner basis for a suitable
term order. This generalizes as follows:

\begin{thm}
\label{cool} Let $T$ be an arbitrary binary rooted tree. Modulo
the trivial invariant $\,q_{00 \cdots 0} - 1$,
\begin{itemize}
\item[(a)] the ideal of the
Jukes-Cantor binary model is generated by polynomials of degree
$2$,
\item[(b)] the ideal of the
Jukes-Cantor DNA model is generated by polynomials of degree $1$,
$2$ and $3$,
\item[(c)] the  ideal of the
Kimura 2-parameter model is generated by polynomials of degree
$1$, $2$, $3$~and~$4$,
\item[(d)] the ideal of the
Kimura 3-parameter model is generated by polynomials of degree
$2$, $3$ and $4$.
\end{itemize}
Each of these generating sets has an explicit combinatorial
description and it is a Gr\"obner basis. \end{thm}

The outline for the paper is as follows.  In Section 2 we review
the Fourier transform technique introduced by Evans and Speed
\cite{ES} for diagonalizing group-based models.  This is  done for
arbitrary finite abelian groups, as in \cite{SSE}, and it reduces
our problem to computing the kernel of a monomial map as in
(\ref{smallmap}).  Section 3 turns rooted trees on $m$ leaves into
unrooted trees on $m+1$ leaves, and it introduces ``friendly
labelings'' on abelian groups.   These labelings are used to
classify the linear model invariants, and to set up a coordinate
system modulo the linear invariants.  This generalizes the
construction in \cite{SF}. In Section 4 we state and prove the
precise form of Theorem \ref{main}. It reduces the construction of
invariants to the case of claw trees $K_{1,m}$. This case is
studied in Section 5.

Section 6 is aimed at computational biologists interested in
experimenting with our invariants. Theorem \ref{cool} is derived
by describing the generating sets explicitly.  Section 7 concerns
the question whether phylogenetics really needs all of the many
invariants in our generating sets. We argue that the answer is
affirmative. We demonstrate by means of an example that
algebraically independent invariants do not suffice to
characterize an evolutionary model, in contrast to what was
suggested in \cite{ES, Ha, SSEW}. Conclusions, algorithmic
questions and open problems appear in Section~8.

\section{A Linear Change of Coordinates}

The Fourier transform provides  a linear change of coordinates
that transforms the irreducible variety of distributions of a
group-based model into a toric variety.  Our presentation in this
section is an exposition of the constructions in  \cite{ES} and
\cite{SSE}.  Experts in combinatorial commutative algebra
\cite{berndbook} will be surprised to encounter toric ideals whose
natural coordinate system is the wrong one: the equations in the
given coordinates $ p_{i_1 i_2 \ldots i_m}$ are very far from
binomial, and the task at hand is to find new coordinates $q_{i_1
i_2 \ldots i_m}$ so that the equations become binomials
$\,q^u-q^v$.

\begin{ex} The Jukes-Cantor binary model for the rooted claw tree $K_{1,3}$
has the parameterization
$$
\begin{matrix}
p_{000} \,\, = \,\, \pi_0 \alpha_0 \beta_0 \gamma_0 + \pi_1
\alpha_1 \beta_1 \gamma_1,  & &
p_{001} \,\, = \,\, \pi_0 \alpha_0 \beta_0 \gamma_1 + \pi_1 \alpha_1 \beta_1 \gamma_0,  \\
p_{010} \,\, = \,\, \pi_0 \alpha_0 \beta_1 \gamma_0 + \pi_1
\alpha_1 \beta_0 \gamma_1,  & &
p_{011} \,\, = \,\, \pi_0 \alpha_0 \beta_1 \gamma_1 + \pi_1 \alpha_1 \beta_0 \gamma_0,  \\
p_{100} \,\, = \,\, \pi_0 \alpha_1 \beta_0 \gamma_0 + \pi_1
\alpha_0 \beta_1 \gamma_1,  & &
p_{101} \,\, = \,\, \pi_0 \alpha_1 \beta_0 \gamma_1 + \pi_1 \alpha_0 \beta_1 \gamma_0,  \\
p_{110} \,\, = \,\, \pi_0 \alpha_1 \beta_1 \gamma_0 + \pi_1
\alpha_0 \beta_0 \gamma_1,  & &
p_{111} \,\, = \,\, \pi_0 \alpha_1 \beta_1 \gamma_1 + \pi_1 \alpha_0 \beta_0 \gamma_0.  \\
\end{matrix}
$$
The Fourier transform gives a linear change of coordinates in  the
parameter space,
\begin{eqnarray*}
& \pi_0 \, = \frac{1}{2} ( r_0 +  r_1),\,\,
  \pi_1 \, = \frac{1}{2} (r_0 - r_1) ,\,\,
\alpha_0 \, = \frac{1}{2} ( a_0 +  a_1),\,\,
\alpha_1 \, = \frac{1}{2} (a_0 - a_1),\\
& \beta_0 \, = \frac{1}{2} ( b_0 +  b_1),\,\, \beta_1 \, =
\frac{1}{2} (b_0 - b_1) ,\,\,
 \gamma_0 \, = \frac{1}{2} ( c_0 +  c_1),\,\,
\gamma_1 \, = \frac{1}{2} (c_0 - c_1),
\end{eqnarray*}
and it simultaneously gives a linear change of coordinates in the
probability space:
\begin{equation}
\label{fourier3}
 p_{ijk} \quad = \quad \sum_{r=0}^1 \sum_{s=0}^1 \sum_{t=0}^1
 (-1)^{ir+js+kt} \cdot  q_{rst}.
 \end{equation}
After these coordinate changes, our model is given by the monomial
parameterization:
$$ \begin{matrix}
q_{000} \,\, = \,\, r_0 a_0 b_0 c_0  ,\,\, & q_{001} \,\, = \,\,
r_1 a_0 b_0 c_1   ,\,\,& q_{010} \,\, = \,\, r_1 a_0 b_1 c_0
,\,\,&
q_{011} \,\, = \,\, r_0 a_0 b_1 c_1  ,\\
q_{100} \,\, = \,\, r_1 a_1 b_0 c_0  ,\,\, & q_{101} \,\, = \,\,
r_0 a_1 b_0 c_1   ,\,\,& q_{110} \,\, = \,\, r_0 a_1 b_1 c_0
,\,\,& q_{111} \,\, = \,\, r_1 a_1 b_1 c_1  .
\end{matrix} $$
The toric ideal of algebraic relations among these monomials has
the following Gr\"obner basis:
$$
\bigl\{ \,q_{001} q_{110} - q_{000} q_{111} \,,\,\,\, q_{010}
q_{101} - q_{000} q_{111} \,,\,\,\, q_{100} q_{011} - q_{000}
q_{111} \, \bigr\}. $$ The  inverse to (\ref{fourier3}) now
translates each of the three binomials into a quadric with eight
terms, e.g.,
$$ p_{001} p_{010} + p_{001} p_{100} - p_{000} p_{011} - p_{000} p_{101}
+ p_{100} p_{111} - p_{101} p_{110} + p_{010} p_{111} - p_{011}
p_{110}. $$ These three eight-term quadrics generate the ideal of
phylogenetic invariants for this model. \qed
\end{ex}

Recall from (\ref{recursive P}) and (\ref{unknownP}) that  the
joint distribution of a Markov model on a tree $T$ has the form
\begin{equation}
\label{evolution1}
 p_{g_1 \ldots g_m} \quad = \quad
 \sum \, \pi_{g_r} \! \prod_{v \in \mathcal{V}(T)
  \setminus \{r\}} A^{(v)}_{g_{a(v)},g_v},
  \end{equation}
  where the sum is over all states of
  the interior nodes of the tree $T$.
  Here we are concerned with the case when
 the states of the random variables are the elements
  of a finite additive abelian group $G$,
  and the transition matrix entry $A^{(v)}_{g_{a(v)},g_v}$
   depends only on the difference
  of $g_{a(v)}$ and $g_v$ in $G$.  We denote this entry by
   $f^{(v)}({g_{a(v)}-g_v})$. Hence group based models
  of evolution have the form
  \begin{equation}
\label{evolution2} p_{g_1,\ldots,g_m} \,\, = \,\,
  p(g_1,\ldots, g_m) \quad = \quad
   \sum \pi(g_r) \prod_{v \in \mathcal{V}(T)
  \setminus \{r\}} f^{(v)}({g_{a(v)}-g_v}).
  \end{equation}
  The right hand side is a polynomial of degree equal to the number of
  edges of $T$ plus one, and the number  of terms of
  this polynomial is $k$ raised to the number of interior nodes of $T$.
  Our aim is to perform a linear change of  coordinates so that this
  big polynomial becomes a monomial.

  The \emph{dual group} to $G$
  (or \emph{character group of $G$})
  is the group of all group homomorphisms from $G$ into the
multiplicative group of complex numbers.  It is denoted by
$\,\widehat{G} = \mathrm{Hom}(G, \cc^{\times})$. The elements of
$\widehat{G}$ are the characters of $G$ and a typical element of
$\widehat{G}$ is denoted by the letter $\chi$.

Given any function $f : G \rightarrow \cc$, the Fourier transform
$\hat{f}$ is the function $\hat{f}: \widehat{G} \rightarrow \cc$
defined by
$$\hat{f}(\chi)  \quad = \quad \sum_{g \in G} \chi(g)f(g). $$
Given two functions $f_1$ and $f_2$ on $G$, their
\emph{convolution} $f_1 * f_2$ is the new function defined by
$$ (f_1 * f_2 )(g) \quad = \quad \sum_{h \in G} f_1(h)f_2(g-h) .$$
The following facts about the dual group and the
 Fourier transform are well-known.

\begin{lemma}\label{lem:fourier}
Let $f_1, f_2$ be functions from a finite abelian group $G$ to
$\cc$ and $\mathbf{1}$ the constant function.
\begin{enumerate}
\item[(a)] The group $G$ and the dual group $ \widehat{G}$ are isomorphic
as abstract groups.
\item[(b)] Fourier transform turns convolution into multiplication, i.e.,
 $\widehat{f_1*f_2} = \widehat{f_1} \cdot \widehat{f_2}$, and
\item[(c)] $\,\widehat{\mathbf{1}}(\chi) = |G| \,$ if $\,\chi = 1 \,$
(the unit in $\widehat{G})$, and $\,\widehat{\mathbf{1}}(\chi) = 0
\,$ otherwise.
\end{enumerate}
\end{lemma}

The main theorem about discrete Fourier analysis and group based
models is that the Fourier transform of the joint distribution has
a parameterization that can be written in product form.  Hence, a
phylogenetic model with group structure is a toric variety.  Note
that if the abelian group for the group model is any group other
than $\mathbb{Z}_2^l$, then the coordinate transformation requires
the complex numbers. Before we state the general result, we
illustrate the idea
 with a small example.

\begin{ex}[Fourier transform for the simplest trees]
\label{ex:base} Let $T = K_{1,m}$ be the tree  whose only nodes
are the $m$ leaves and the root. The joint probability of a group
based model is given by
\begin{equation}
\label{onlyroot} p(g_1,g_2, \ldots, g_m) \quad = \quad \sum_{h \in
G}\pi(h) \prod_{i=1}^m f^{(i)}(h-g_i).
\end{equation}
  We will take the Fourier transform of this probability
density with respect to the group $G^m$.  To do this, we replace
the root distribution $\,\pi : G \rightarrow \rr \,$ by a new
function $\,\widetilde{\pi} : G^m \rightarrow \rr\,$ as follows:
$$ \widetilde{\pi}(h_1, \ldots, h_m) \quad  = \quad \left\{ \begin{array}{cl}
\pi(h_1) & \mbox{ if } h_1 = h_2 = \cdots = h_m \\
0 & \mbox{  otherwise} \end{array} \right.$$
 Then we have
$$p(g_1,g_2, \ldots, g_m) \quad = \quad \sum_{(h_1, \ldots, h_m) \in G^m} \!\!\!
\tilde{\pi}(h_1,\ldots, h_m) \prod_{i=1}^m f^{(i)}(h_i-g_i).$$
Thus $p$ is the convolution of two functions on $G^m$. Taking the
Fourier transform yields
$$ q(\chi_1, \ldots, \chi_m) \quad = \quad \widehat{\widetilde{\pi}}
(\chi_1, \ldots, \chi_m) \prod_{i=1}^m \widehat{f^{(i)}}(\chi_i)$$
by the convolution formula and the independence of the $f^{(i)}$
in the Fourier transform.  Furthermore,
\begin{eqnarray*}
\widehat{\widetilde{\pi}}(\chi_1, \ldots, \chi_m) & \,\, =
\,\,\,\,\sum_{(g_1, \ldots, g_m)
  \in G^m} \left<(\chi_1, \ldots, \chi_m), (g_1, \ldots, g_m)\right>
  \cdot \widetilde{\pi}(g_1, \ldots, g_m) \\
& = \quad \sum_{g \in G} \left< \chi_1 \chi_2 \cdots \chi_m
,g\right> \cdot \pi(g) \quad =\quad
  \widehat{\pi}(\chi_1 \chi_2 \cdots \chi_m),
  \end{eqnarray*}
 and hence
\begin{equation}
\label{simplesttrees} q(\chi_1, \ldots, \chi_m) \quad = \quad
\widehat{\pi}(\chi_1 \cdots \chi_m) \prod_{i=1}^m
\widehat{f^{(i)}}(\chi_i)
\end{equation}
\end{ex}

Example \ref{ex:base} is the base case in the induction needed to
prove the following general result.

\begin{thm} \label{evansspeed} {\rm (Evans-Speed \cite{ES})}
Let $p(g_1, \ldots, g_m)$ be the joint distribution of a group
based model for the phylogenetic tree $T$, parametrized as in
(\ref{evolution2}). Then the Fourier transform of $p$ has the form
\begin{equation}
\label{evolution3} q(\chi_1, \ldots, \chi_m) \quad = \quad
\widehat{\pi}(\chi_1 \cdots \chi_m) \,\cdot \!\!\!\prod_{v \in
\mathcal{V}(T)\setminus \{r\}} \!\!\widehat{f^{(v)}} \bigl( \!
\prod_{l  \in \Lambda(v)} \chi_l \bigr)
\end{equation}
 where $\Lambda(v)$ is the set of leaves which have
$v$ as a common ancestor.
\end{thm}

We refer to \cite{ES} and \cite{SSE} for the proof of Theorem
\ref{evansspeed}. The transformation from (\ref{onlyroot}) to
(\ref{simplesttrees}) is a special case of the transformation from
(\ref{evolution2}) to (\ref{evolution3}). Formula
(\ref{evolution2}) is a polynomial parameterization of the
evolutionary model, and formula (\ref{evolution3}) is a monomial
parameterization of the same model. Since $G$ and $\widehat{G}$
are isomorphic groups, we can rewrite the monomial
parameterization (\ref{evolution3}) as follows:
\begin{equation}
\label{evolution4} q_{g_1, \ldots, g_m} \quad \mapsto \quad
\widehat{\pi}(g_1 +\cdots + g_m) \prod_{v \in
\mathcal{V}(T)\setminus \{r\}}\widehat{f^{(v)}}( \sum_{l
  \in \Lambda(v)} g_l).
 \end{equation}
 We regard this formula as a monomial map from a polynomial ring
 in $|G|^m$ unknowns
 $$ q_{g_1,\ldots,g_m} \, = \, q(g_1, \ldots, g_m) $$
 to the polynomial ring in the
 (not necessarily distinct) unknowns $\widehat{\pi}(g)$
 and $ \widehat{f^{(v)}}(g)$, which are indexed by  nodes of $T$
 and elements  of $G$.  Our aim is to determine the kernel of this map.

\section{Edge Labelings and Linear Invariants}

 In this section, we determine all linear forms
 $\, q_{g_1,\ldots,g_m} - q_{h_1,\ldots,h_m}\,$
 in the kernel of our monomial map (\ref{evolution4}),
 and we set up a convenient coordinate system
 for working modulo these
 \emph{linear invariants}.
Our construction is inspired by the work of Steel and Fu \cite{SF}
on classifying the linear invariants.

  We first add an extra edge
at the root of $T$ to achieve a new tree with $m+1$ leaves. To
keep notation simple, we denote the new tree also by $T$. Let
$E(T)$ be its set of edges. We next associate a set of parameters
to each $e \in E(T)$ by ``moving'' the parameters from a given
node to the edge directly above it.  Given an assignment of group
elements $(g_1, \ldots, g_m)$ to the $m$ leaves of $T$, we get,
for each edge $e$ of $T$ an assignment of a group element $g(e)$
as follows:
$$ g(e) \quad  = \quad \sum_{v \in \Lambda(e)} g_v .$$
Here $\Lambda(e)$ is the set of leaves below $e$. With this
notation, we have eliminated the special distinction of the root
distribution, and our monomial parameterization (\ref{evolution4})
can be rewritten as
\begin{equation}
\label{evolution5}
 q_{g_1 \ldots g_m} \,\,\, \mapsto \prod_{e \in E(T)} f^{(e)}(g(e)).
 \end{equation}
If the unknowns $\,f^{(e)}(g)\,$ are all distinct then there are
no linear invariants. This happens in the Jukes-Cantor binary
model and in the Kimura 3-parameter model. However, in general, we
allow the possibility that $f^{(e)}(g)  = f^{(e)}(g')$ for
distinct group elements $g,g' \in G$. To deal with this issue, we
introduce labeling functions. Let $\mathcal{L}$ be a finite set of
labels. A \emph{labeling function} is any function
$$L \,\, : \,\, G \rightarrow \mathcal{L}  $$
such that $f^{(e)}(g)  = f^{(e)}(g')$ if and only if $L(g)  =
L(g')$. For the time being, we will assume that the labeling
function associated to each edge of the tree is the same for every
edge.  However, we will show later that this assumption can be
dropped in some special instances.

 Given such a labeling function $L$ we can now write our monomial
parameterization (\ref{evolution5}) in a standard commutative
algebra notation.
  For every edge $e$ of the tree $T$
and every label $l \in \mathcal{L}$ we introduce an indeterminate
$a^{(e)}_l$. These indeterminates are now distinct. The polynomial
ring in these unknowns with complex coefficients is denoted
 $\,\cc[a^{(e)}_l]$.  Similarly, $\cc[q_{g_1 \ldots g_m}]$ is the
 polynomial ring generated by the Fourier coordinates.
 We wish to study the ring homomorphism
\begin{equation}
\label{evolution6} \cc[q_{g_1 \ldots g_m}] \rightarrow
\cc[a^{(e)}_l] \, ,\,\, q_{g_1 \ldots g_m} \mapsto \prod_{e \in
E(T)}
 a^{(e)}_{L(g(e))}.
 \end{equation}
 The kernel of this map is the toric ideal of
  phylogenetic invariants in the Fourier transform of the
probabilities.  We denote this ideal by $I_{T,L}$ suppressing
dependence on the group $G$. From this description, we immediately
can deduce the structure of the linear phylogenetic invariants.

\begin{prop}[Linear Invariants]
\label{lininv} The vector space of linear polynomials in the ideal
$I_{T,L}$ is spanned by all differences  $\, q_{g_1 \ldots g_m} -
q_{h_1 \ldots  h_m}\,$ where $L(g(e)) = L(h(e))$ for all edges $e$
of $T$.
\end{prop}

\begin{proof}
Since $I_{T,L}$ is a toric ideal, it has a vector space basis
consisting of binomials $q^u-q^v$. In particular, the subspace of
linear polynomials in $I_{T,L}$ is spanned by differences of
unknowns $\, q_{g_1 \ldots  g_m} - q_{h_1 \ldots  h_m}$. Such a
difference lies in $I_{T,L}$ if and only if $\, \prod_{e \in E(T)}
a^{(e)}_{L(g(e))} \, =\, \prod_{e \in E(T)} a^{(e)}_{L(h(e))}$.
Since the unknowns $a^{(e)}_l$ are all distinct, this happens if
and only if $\, L(g(e)) \,  = \, \sum_{v \in \Lambda(e)} g_v \,$
coincides with $\, L(h(e)) \,  = \, \sum_{v \in \Lambda(e)} h_v
\,$ for all $e \in T$.
\end{proof}

We now introduce coordinates for the polynomial ring $\cc[q_{g_1
\ldots g_m}]$ modulo the ideal generated by the linear invariants
in $I_{T,L}$. The labeling function $L: G \rightarrow \mathcal{L}$
induces the function
\begin{equation}
\label{tolabels} L^T :G^m \rightarrow \mathcal{L}^{E(T)} \,,\,\,
(g_1,\ldots,g_m) \mapsto \bigl( L(g(e)) \bigr)_{e \in E(T)}.
\end{equation}
Let ${\rm im}(L^T)$ denote the image of this map. We call ${\rm
im}(L^T)$ the set of \emph{consistent labelings} of the tree $T$.
For each $\lambda \in {\rm im}(L^T)$ we introduce a new unknown
$q_\lambda$. These generate a new polynomial ring.

 Proposition \ref{lininv} implies that our monomial map (\ref{evolution6})
is the composition of the map
$$ \cc \bigl[\,q_{g_1 \ldots g_m} \,:\,
(g_1,\ldots,g_m) \in G^m\, \bigr] \rightarrow \cc \bigl[\,
q_\lambda \, :\, \lambda \in {\rm im}(L^T)\, \bigr] \,,\,\, q_{g_1
\ldots g_m} \mapsto q_{L^T(g_1,\ldots,g_m)}, $$ and the following
monomial map which has no linear forms in its kernel:
\begin{equation}
\label{evolution7}
 \cc \bigl[ \,q_\lambda \, :\, \lambda \in {\rm im}(L^T) \, \bigr]
 \rightarrow
 \cc \bigl[\, a^{(e)}_l \, :\, e \in E(T) ,\, l \in \mathcal{L} \,\bigr]\,,\,
 \, q_\lambda \, \, \mapsto  \prod_{e \in E(T)} a^{(e)}_{\lambda(e)} .
 \end{equation}
 Our objective is to determine the kernel of  the monomial map (\ref{evolution7}).
This kernel is the toric ideal $I_{T,L}$ modulo linear invariants.
We use the same symbol $I_{T,L}$ to denote the kernel of
(\ref{evolution7}).

 Our main result, which will be stated and proved in the next section,
 is valid only for a certain subclass of labeling functions. These will be called
 the friendly labeling functions.
Fortunately, all labeling functions which arise naturally in
statistical models of evolution are friendly.

\begin{defn}
Fix a labeling function $L: G \rightarrow \mathcal{L}$ on the
group $G$. For $m \geq 3$ consider the set
$$Z \quad = \quad  \bigl\{ (g_1, \ldots, g_m) \in G^m \,:\,
\sum_{i=1}^{m-1} g_i = g_m \bigr\}.$$

\noindent Consider the induced map $\widetilde{L}:Z \subset G^m
\rightarrow \mathcal{L}^{m}$ and denote by $\pi_i$ the projection
$\pi_i:G^m \rightarrow G$ onto the $i$-th coordinate. The function
$L$ is called $m$-\emph{friendly} if, for every $l=(l_1, \ldots,
l_m) \in \widetilde{L}(Z) \subset \mathcal{L}^m$,
\begin{equation}
\label{friendly} \pi_i(\widetilde{L}^{-1}(l))\,= \, L^{-1}(l_i)
\qquad \qquad \hbox{for all}  \,\, i = 1, \ldots, m.
\end{equation}
Note that the inclusion ``$\subseteq$'' always holds. But for most
labeling functions it will be strict.  Note that  $Z$ is the set
of all allowable assignments of group elements to the edges of the
unrooted tree $T= K_{1,m}$.  The definition of $m$-friendly
guarantees that if a particular labeling $\lambda$ comes from an
assignment of group elements, then any choice of a group element
to one particular edge $e$ which is consistent with $\lambda$ at
$e$ can be extended to an assignment that is consistent with
$\lambda$ on all the edge of $K_{1,m}$.

\begin{ex}
Let $G = \mathbb{Z}_4$ and $\mathcal{L} = \{0,1,2\}$.  Then the
labeling function $L$ defined by
$$L(0)=0,\,\, L(1)=1, \,\, L(2) = L(3)= 2$$
 is not $3$-friendly because
 $L^{-1}(2) = \{2,3\}$ strictly contains
 $\,\pi_3( \widetilde{L}^{-1}((1,1,2)))= \pi_3( \{(1,1,2)\})
 =  \{2\}  $.
\end{ex}

The next example looks similar, but it is, in fact, much more
friendly.

\begin{ex} [The Kimura 2-parameter labeling function]
\label{kimuralabeling} Let $\,G = \mathbb{Z}_2 \times \mathbb{Z}_2
$ and  $\mathcal{L} = \{0,1,2\}$. The Kimura 2-parameter model
corresponds to
 the labeling function $L$ defined by
$$ L((0,0))= 0, \,\, L((0,1)) = 1, \,\, L((1,0))= L((1,1))=2 .$$
It can be checked by an explicit calculation that $L$ is
$3$-friendly.
\end{ex}

We say that a labeling function $L : G \rightarrow \mathcal{L}$ is
\emph{friendly} if it is $m$-friendly for all $m \geq 3$.
\end{defn}

\begin{lemma}\label{lem:3fr}
Labeling functions that are $3$-friendly are friendly.
\end{lemma}

\begin{proof}
We will show that a labeling function that is $3$-friendly and
$m$-friendly is also $(m+1)$-friendly.  Let $l \in
\widetilde{L}(Z)$. We will show that
$\pi_{m+1}(\widetilde{L}^{-1}(l)) = L^{-1}(l_{m+1})$. Let $l' =
(L(g_1 + g_2), L(g_3), \ldots, L(g_{m+1}))$ where $(g_1, \ldots,
g_{m+1}) \in \widetilde{L}^{-1}(l)$.  Since $L$ is $m$-friendly,
for every $h_{m+1} \in L^{-1}(l_{m+1})$ there is an assignment of
group elements $h'= (h_2', h_3, \ldots, h_{m+1})$. Furthermore,
$L$ is $3$-friendly so there is some choice of group assignment
$(h_1,h_2, h_2')$ that realizes the labeling
$(L(g_1),L(g_2),L(g_1+g_2))$.  But then $h = (h_1, h_2, h_3,
\ldots, h_{m+1})$ has $\pi_{m+1}(h) = h_{m+1}$ as desired.
\end{proof}

Lemma \ref{lem:3fr} says that checking whether a labeling is
friendly can be done simply with a finite computation.  The point
of studying friendly labelings is that consistent labelings
``glue'' together.  We will now make this statement explicit.  Let
$e$ be an interior edge of the tree $T$.  Denote by $T_{e,-}$ the
tree obtained from $T$ by taking the edge $e$ and all the edges
below $e$.  Denote by $T_{e,+}$ the tree obtained from $T$ be
taking the edge $e$ and all edges not in $T_{e,-}$.  Then we have
the following

\begin{lemma}\label{lem:labglue}
Let $\lambda^-$ and $\lambda^+$ be consistent labelings of
$T_{e,-}$ and $T_{e,+}$ respectively, i.e.~$\lambda^- \in {\rm
im}(L^{T_{e,-}})$ and $\lambda^+ \in {\rm im}(L^{T_{e,+}})$.
Suppose furthermore that $\lambda^-(e) = \lambda^+(e)$. Then the
labeling $\lambda$ of $T$ obtained from $\lambda^-$ and
$\lambda^+$ by labeling edges of $T$ appropriately is consistent,
i.e., $\lambda \in {\rm im}(L^T)$.
\end{lemma}

\begin{proof}
Since $\lambda^+$ and $\lambda^-$ are consistent, there is some
assignment of group elements to the edges of $T_{e,+}$ and
$T_{e,-}$ that comes from $(L^{T_{e,+}})^{-1}({\rm
im}(L^{T_{e,+}}))$ and $(L^{T_{e,-}})^{-1}({\rm
im}(L^{T_{e,-}}))$.   We will now construct an assignment of group
elements of the edges of $T$ that belongs to $(L^T)^{-1}({\rm
im}(L^T))$.  First take any assignment which is compatible with
$\lambda^+$ on $T_{e,+}$.  This assigns some group element to the
edge $e$.  Let $v$ be the nonleaf vertex of $T_{e,-}$ incident to
$e$. Since $L$ is friendly, and $\lambda^-$ is consistent, there
exists an assignment of group elements to all the other edges
incident to $v$ which is compatible with $\lambda^-(e)$ and is
locally consistent.  By induction on the number of interior
vertices of $T_{e,-}$ we construct a globally consistent
assignment of group elements to the edges of $T$.
\end{proof}

Lemma \ref{lem:labglue} is the main technical result upon which
all our combinatorial constructions of generators and Gr\"obner
bases rest.  Indeed, as we will see, it implies that phylogenetic
invariants of group based models with friendly labelings are only
determined by local features of the tree. We  conclude this
section with some examples of friendly labeling functions.

\begin{ex} \label{injectivelabeling}
Let $G$ be any finite abelian group.  Any function $L:G
\rightarrow \mathcal{L}$ that is injective is friendly for trivial
reasons (the two sets in (\ref{friendly}) are singletons and hence
equal). For similar reasons, if $\mathcal{L}$ consists of elements
of a group and $L$ is a group homomorphism then $L$ is friendly.
\end{ex}

\begin{ex}[The Jukes-Cantor labeling function]
\label{jukeslabeling} Let $\mathcal{L} = \{0,1\}$ and $L$ the
function
$$L(g) = \left\{ \begin{array}{cl}
0 & \mbox{  if } g = 0 \\
1 & \mbox{  otherwise} \end{array} \right.  $$ Then $L$ is
friendly for any group $G$. It corresponds to the Jukes-Cantor
models when $\,G = \mathbb{Z}_2^r$.
\end{ex}

\begin{ex}
The  Kimura 2-parameter labeling function of Example
\ref{kimuralabeling} is friendly by  Lemma \ref{lem:3fr}.
 \end{ex}

\section{The Main Result}

We will now state and prove our main result concerning
 the ideal of phylogenetic invariants
of any group based model with friendly labeling function $L$.
 We consider the toric ideal $I_{T,L}$ which is the kernel
 of the monomial map (\ref{evolution7}), and we construct
 minimal generators and a Gr\"obner basis for $I_{T,L}$ out of purely
 local information in the tree. This Gr\"obner basis
 is a list of binomials $q^u - q^v$ in the unknowns
 $q_l$ which are indexed by the consistent  labelings $l \in {\rm im}(L^T)$.
 In order to transform the binomials into polynomials in the
 probabilities $p_{g_1,\ldots,g_m}$, one must reverse
 the transformations described in Sections 2 and 3.
 In Section 6, we will
 characterize the consistent
labelings and examine the relevant transformations for the four
standard models of Theorem \ref{cool}. Throughout this section, we
assume that $\,L :  G \rightarrow \mathcal{L} \,$ is an arbitrary
friendly labeling on a finite abelian group $G$.

For ease of notation, we write the monomials in the unknowns $q_l$
using the \emph{tableau notation}.  This means that any monomial
$\,M = q_{l_1}q_{l_2} \cdots q_{l_d}\,$ is written as a matrix of
format $d \times |E(T)|$:
$$ M \quad = \quad  \left[ \begin{array}{c}
l_1 \\
l_2 \\
\vdots \\
l_d  \end{array} \right].$$ Such a matrix with entries in $L$ is
called a \emph{tableau}. The columns of a tableau are indexed by
the edges of the tree $T$ under consideration. The number of rows
of $M$ is the degree $d$ of the monomial. Two tableaux represent
the same monomial if they are related by a permutation of rows.

  Binomials $q^u-q^v$ in the unknowns $q_l$ are
represented as formal differences $M-M'$ of tableaux.  Notice that
it is easy to check whether a given binomial $ M - M'$ lies in the
toric ideal $I_{T,L}$.

\begin{rmk}
\label{membership} Let $M$ and $M'$ be two tableaux of format $d
\times |E(T)|$ with entries in $L$. Then the binomial $M-M'$ lies
in the ideal $I_{T,L}$ if and only if the following two conditions
hold:
\begin{enumerate}
\item[(a)] each row of $M$ and each row of $M'$ is a consistent labeling for
the tree $T$, and
\item[(b)] for each edge $e \in E(T)$, the
 multiset of labels in column $e$ is the same in $M$ and in $M'$.
 \end{enumerate}
\end{rmk}

 We are now ready
to construct the binomials that will constitute the Gr\"obner
bases of $I_{T,L}$. Let $e$ be an interior edge of $T$, and let
$T_{e,-}$ and $T_{e,+}$ be the two subtrees as in Lemma
\ref{lem:labglue}. After relabeling the edges of $T$, every
tableau $M$ can be written in three groups of columns,
$$ M \quad = \quad  \left[ \begin{array}{ccc}
l_1 & m_1 & n_1\\
l_2 & m_2 & n_2\\
\vdots & \vdots & \vdots \\
l_d & m_d & n_d \end{array} \right] ,$$ where the left columns
(with entries $l_i$) correspond to the edges in $T_{e,-}
\backslash \{e\}$, the middle column corresponds to the edge $e$,
and the right columns correspond to the edges in $T_{e,+}
\backslash \{e\}$.

\begin{lemma}\label{lem:ind}
Let $(l_1, m ,n_1)$ and $(l_2, m , n_2)$ be consistent labelings
of $T$. Then the quadratic binomial
$$g \quad = \quad \left[ \begin{array}{ccc}
l_1 & m & n_1 \\
l_2 & m & n_2 \end{array} \right] - \left[ \begin{array}{ccc}
l_1 & m & n_2 \\
l_2 & m & n_1 \end{array} \right] $$ lies in the toric ideal
$I_{T,L}$.
\end{lemma}

\begin{proof}
The labelings $(l_1,m)$ and $(l_2,m)$ are consistent for  the
subtree $T_{e,-}$, and the labelings $(m,n_1)$ and $(m,n_2)$ are
consistent for the subtree $T_{e,+}$. By Lemma \ref{lem:labglue},
the labelings $(l_1,m,n_2)$ and \ $(l_2,m,n_1)$ are consistent for
the big tree $T$. Remark \ref{membership} implies that $\,g \in
I_{T,L}$.
\end{proof}

\begin{defn}
Denote by ${\rm Quad}(e,T)$ the set of all the quadratic binomials
$g$ from Lemma \ref{lem:ind}.
\end{defn}

Consider now an arbitrary binomial in the ideal $I_{T,L}$. It has
the form
$$ h \quad  = \quad \left[ \begin{array}{ccc}
l_1 & m_1 & n_1 \\
\vdots & \vdots & \vdots \\
l_d & m_d & n_d \end{array} \right] - \left[ \begin{array}{ccc}
l'_1 & m'_1 & n'_1 \\
\vdots & \vdots & \vdots \\
l'_d & m'_d & n'_d \end{array} \right]  $$
 where the $m_i$ and $m'_i$ are single labels
corresponding to the edge $e$, the $l_i$ and $l'_i$ are consistent
labelings of $T_{e,-} \backslash \{e\}$, and the $n_i$ and $n'_i$
are consistent labelings of
 $T_{e^+} \backslash \{e\} $.  Note that since the binomial $h$ belongs to $I_{T,f}$,
the multiset of labels which appears on the edge $e$ must be the
same for both terms of $h$.  Hence, after rearranging the rows of
the tableau we may write
$$ h  \quad = \quad  \left[ \begin{array}{ccc}
l_1 & m_1 & n_1 \\
\vdots & \vdots & \vdots \\
l_d & m_d & n_d \end{array} \right] - \left[ \begin{array}{ccc}
l'_1 & m_1 & n'_1 \\
\vdots & \vdots & \vdots \\
l'_d & m_d & n'_d \end{array} \right].  $$

Every binomial in $I_{T,L}$ restricts to a binomial in
$I_{T_{e,-},L}$ and to a binomial in $I_{T_{e,+},L}$. Namely, if
$h$ is the binomial above, then the following binomial lies in
$I_{T_{e,-},L}$ :
$$ h|_{T_{e,-}} \quad  = \quad  \left[ \begin{array}{cc}
l_1 & m_1  \\
\vdots & \vdots  \\
l_d & m_d  \end{array} \right] - \left[ \begin{array}{cc}
l'_1 & m_1  \\
\vdots & \vdots  \\
l'_d & m_d  \end{array} \right]  $$
 Similarly, deleting the left columns yields a binomial
$h|_{T_{e,+}}$  in $I_{T_{e,+},L}$. We now state  a constructive
converse, from which binomials in $I_{T_{e,-},L}$ and
$I_{T_{e,+},L}$ can be extended to binomials in $I_{T,L}$.

\begin{lemma}\label{lem:newext}
Let $g$ be a binomial in $I_{T_{e,-},L}$ written in tableau
notation as
 $$ g \quad = \quad \left[ \begin{array}{cc}
l_1 & m_1  \\
\vdots & \vdots  \\
l_d & m_d  \end{array} \right] - \left[ \begin{array}{cc}
l'_1 & m_1  \\
\vdots & \vdots  \\
l'_d & m_d  \end{array} \right].  $$ Let $n_1, \ldots, n_d$ be
sequences of labels such that each  $(m_i,n_i) $ is a consistent
labeling of $T_{e,+}$. Then
$$ g^* = \left[ \begin{array}{ccc}
l_1 & m_1 & n_1 \\
\vdots & \vdots & \vdots \\
l_d & m_d & n_d \end{array} \right] - \left[ \begin{array}{ccc}
l'_1 & m_1 & n_1 \\
\vdots & \vdots & \vdots \\
l'_d & m_d & n_d \end{array} \right]  $$ is a binomial in
$I_{T,L}$.
\end{lemma}

\begin{proof}
Restricting the two tableaux to the tree $T_{e,-}$ and $T_{e,+}$
shows that the multiset of labels which appears on each edge are
the same. In fact, we have
$$ g^*|_{T_{e,-}} = g \quad \hbox{and}
\quad g^*|_{T_{e,+}} = 0. $$
  We must  check that each of  $(l_i,m_i,n_i)$ and $(l'_i,m_i,n_i)$ is a
consistent labeling on $T$. Lemma  \ref{lem:labglue} implies this
because
 $(l_i,m_i) $ and $(l_i',m_i)$ are
consistent on $T_{e,-}$ and $(m_i,n_i)$ is consistent on
$T_{e,+}$.
\end{proof}

\begin{defn}
Let $\mathcal{B} $ be a collection of binomials in
$\,I_{T_{e.-},L}$. We define
 $\, {\rm Ext}(\mathcal{B}\rightarrow T)\, $
 to be the set of all binomials $g^*$
 where $ g$ ranges over $\mathcal{B}$ and
 $n_1,\ldots,n_d$ ranges over of sequences of labels as
 in Lemma \ref{lem:newext}.
 Similarly, we define $\,{\rm Ext}(T \leftarrow \mathcal{B})\,$
 for any collection of binomials $\,\mathcal{B}$ in $ \, I_{T_{e,+},L}$.
\end{defn}

The first main result of this paper is the following theorem.

\begin{thm}\label{thm:gromain}
Let $T$ be any tree with a friendly labeling $L: G \rightarrow
\mathcal{L}$ and let $ e$ be any interior edge of $T$. Suppose
that $\mathcal{B}_-$ is a binomial generating set for
$I_{T_{e,-},L}$ and let $\mathcal{B}_+$ be a binomial generating
set for $I_{T_{e,+},L}$.  Then the following set of binomials
generates the toric ideal $I_{T,L}$:
\begin{equation}
\label{WeLikeThisSet}
 {\rm Ext}(\mathcal{B}_- \rightarrow T)
 \,\, \cup \,\,{\rm Ext}(T \leftarrow \mathcal{B}_+)
\,\, \cup \,\, {\rm Quad}(e,T) .
\end{equation}
Moreover, if $\mathcal{B}_-$ is a Gr\"obner basis for
$I_{T_{e,-},L}$ and $\mathcal{B}_+$ is a Gr\"obner basis for
$I_{T_{e,+},L}$, then there exists a term order on $\, \cc \bigl[
\,q_\lambda \, :\, \lambda \in {\rm im}(L^T) \, \bigr]\,$ such
that the set in (\ref{WeLikeThisSet}) is a Gr\"obner basis for
$I_{T,L}$.
\end{thm}

\begin{proof}
We first prove the second statement concerning Gr\"obner bases. To
this end we need to specify the term orders. Let $\prec_-$ be any
term order on $\, \cc \bigl[ \,q_\lambda \, :\, \lambda \in {\rm
im}(L^{T_{e,-}}) \, \bigr]\,$ such that $\mathcal{B}_-$ is a
Gr\"obner basis for $I_{T_{e,-},L}$ and $\prec_+$ any term  order
on $\, \cc \bigl[ \,q_\lambda \, :\, \lambda \in {\rm
im}(L^{T_{e,+}}) \, \bigr]\,$ such that $\mathcal{B}_+$ is a
Gr\"obner basis for $I_{T_{e,+},L}$. Finally, let us define a
reverse lexicographic term order $\prec_Q$ which makes
$\mathrm{Quad}(e,T)$ a Gr\"obner basis for the ideal it generates.
We do this by first taking any total order $\prec_1$ on the labels
of the edge $e$, then taking total orders $\prec_2$ on ${\rm
im}(L^{T_{e,-}})$ and
 $\prec_3$ on  ${\rm im}(L^{T_{e,+}})$  which are
refinements of $\prec_1$.  The revlex term order $\prec_Q$ is
obtained by declaring $q_{\lambda_1} \prec_Q q_{\lambda_2}$ if and
only if
$$ \lambda_1^- \prec_2 \lambda_2^- \quad \hbox{or} \quad
( \lambda_1^- = \lambda_2^- \,\,\, \hbox{and} \,\,\, \lambda_1^+
\prec_3 \lambda_2^+). $$
 We construct a  product term
order $\prec_T$ on the polynomial ring $\, \cc  \bigl[ \,q_\lambda
\, :\, \lambda \in {\rm im}(L^T) \, \bigr]\,$ as follows. If  $M$
and $M'$ are monomials (tableaux with columns indexed by $E(T)$)
then $\, M \prec_T M' \,$ if and only if
\begin{enumerate}
\item $ M|_{T_{e,-}} \prec_- M'|_{T_{e,-}}$, or
\item $ M|_{T_{e,-}}  =  M'|_{T_{e,-}}$ and $M|_{T_{e,+}} \prec_+
  M'|_{T_{e,+}}$, or
\item $M|_{T_{e,-}} =  M'|_{T_{e,-}}$ and $M|_{T_{e,+}} =
  M'|_{T_{e,+}}$ and $M \prec_Q M'$.
\end{enumerate}
Our goal is to show that the set
 (\ref{WeLikeThisSet}) is a Gr\"obner basis for $I_{T,L}$
 with respect to the term order $\prec_T$, i.e.,
  the leading term of every binomial $g$ in $I_{T,L}$ is
divisible by the leading term of some binomial from
(\ref{WeLikeThisSet}). To prove this, we consider an arbitrary
binomial in our toric ideal:
$$g \,\,\,= \,\,\,M' - M \,\,\, = \,\,\, \left[ \begin{array}{ccc}
l_1 & m_1 & n_1 \\
\vdots & \vdots & \vdots \\
l_d & m_d & n_d \end{array} \right] - \left[ \begin{array}{ccc}
l'_1 & m_1 & n'_1 \\
\vdots & \vdots & \vdots \\
l'_d & m_d & n'_d \end{array} \right] \quad \in \,\, I_{T,L}. $$
Suppose that $M'$ is the leading term of $g$.
 There are precisely three different ways this can happen,
  according to the three cases in the definition of
  $\prec_T$.  Each case will be analyzed separately.

\emph{Case 1}:  Suppose that  $ M|_{T_{e,-}} \prec_-
M'|_{T_{e,-}}$. Then $g|_{T_{e,-}}$ is a nonzero binomial in
$I_{T_{e,-},L}$ and $M'|_{T_{e,-}}$ is its leading term.  Since
$\mathcal{B}_-$ is a Gr\"obner basis  there exists a binomial $h=
N' - N \in \mathcal{B}_-$ whose leading term $N'$  divides $M'$.
Upon reordering the rows of $M'$ and $M$, we may suppose that
$$ h \,\,\,= \,\,\,N' - N \quad = \quad \left[ \begin{array}{cc}
l_1 & m_1  \\
\vdots & \vdots  \\
l_i & m_i  \end{array} \right] - \left[ \begin{array}{cc}
l''_1 & m_1  \\
\vdots & \vdots  \\
l''_i & m_i  \end{array} \right] \qquad \hbox{for some} \,\, i
\leq d . $$ Here  $l_1, \ldots, l_i$ and $m_1, \ldots, m_i$ are
the same labels that appear in $M'$.  Now we consider the binomial
$\,h^* \in {\rm Ext}(\mathcal{B}_- \rightarrow T)\,$ obtained
 by appending the labels $n_1, \ldots, n_i$:
$$ h^* \,\,\,= \,\,\,(N')^* - N^* \,\,\,= \,\,\,\left[ \begin{array}{ccc}
l_1 & m_1 & n_1 \\
\vdots & \vdots & \vdots \\
l_i & m_i & n_i \end{array} \right] - \left[ \begin{array}{ccc}
l''_1 & m_1 & n_1 \\
\vdots & \vdots & \vdots \\
l''_i & m_i & n_i \end{array} \right].  $$ The tableau $(N')^*$ is
the leading term of $h^*$  with respect to $\prec_T$, and $(N')^*$
divides $M'$ as desired.

\emph{Case 2}:  Suppose $M|_{T_{e,-}} = M|'_{T_{e,-}}$ and
$M|_{T_{e,+}} \prec_+ M'|_{T_{e,+}}$.  Then by the same argument
as in Case 1, we deduce that there is a binomial $h^* \in Ext(T
\leftarrow B_+)$ whose leading term divides $M'$.

\emph{Case 3}:  Suppose that $M|_{T_{e,-}} = M'|_{T_{e,-}}$ and
$M|_{T_{e,+}} = M'|_{T_{e,+}}$ and $M \prec_Q M'$. The only way
that this could happen is if there exists a pair of rows in $M'$,
$(l_1,m,n_1)$ and $(l_2, m, n_2)$, such that $(l_1,m) \prec_1
(l_2,m)$ and $(m,n_1) \succ_2 (m,n_2)$.   But then the binomial $h
\in {\rm
  Quad}(e,T)$ given by
$$ h \,\,\, = \,\,\, N' - N \,\,\,= \,\,\,\left[ \begin{array}{ccc}
l_1 & m & n_1 \\
l_2 & m & n_2 \end{array} \right] - \left[ \begin{array}{ccc}
l_1 & m & n_2 \\
l_2 & m & n_1 \end{array} \right] $$
 has leading term $N'$, and this leading term divides
 the leading term  $M'$ of the binomial $g$.

 These three cases together establish
the second statement: the set (\ref{WeLikeThisSet}) is a Gr\"obner
basis for $I_{T,L}$. Furthermore, for any Gr\"obner bases
$\mathcal{B}_-$ and $\mathcal{B}_+$, we have the equality of
ideals
$$ I_{T,L} \quad = \quad \left<\,{\rm Ext}(\mathcal{B}_- \rightarrow T)\,\right>
 \,\, +  \,\, \left<\,{\rm Ext}(T \leftarrow \mathcal{B}_+) \,\right>
\,\, +  \,\, \left<\, {\rm Quad}(e,T)\, \right> . $$ In this
equation, we may replace ${\rm Ext}( \mathcal{B}_- \rightarrow T)$
with any set  that generates $\,\left< \,{\rm Ext}(\mathcal{B}_-
\rightarrow T)\,\right>$.   But  ${\rm Ext}(
\mathcal{C}_-\rightarrow T)$ generates  $\left<{\rm
Ext}(\mathcal{B}_- \rightarrow T)\right>$ whenever $\mathcal{C}_-$
is a generating set for $I_{T_{e,-},L}$.  A similar statement
holds for $I_{T_{e,+},L}$. This completes the proof of the first
statement in Theorem \ref{thm:gromain}.
\end{proof}

Theorem 1 in the Introduction says that all invariants are
determined by local features of the tree. We shall now state this
result more precisely and derive it as a corollary from Theorem
\ref{thm:gromain}.

 Let $v$ be an interior vertex of the tree $T$, and
  let $e_1, \ldots, e_c$ the edges of $T$ incident to
$v$.  Denote by $T_{v,e_i}$ the subtree $T_{e_i,-}$ or $T_{e_i,+}$
which has $v$ as a leaf.  Given a particular label $l$ for the
edge $e_i$, denote by ${\rm im} (L^{T_{v,e_i}},l)$ the set of all
consistent labelings of $T_{v,e_i}$ which has the label of $e_i$
equal to $l$.  Denote by $T_v$ the subtree of $T$ with only
interior node $v$ and edges $e_1, \ldots, e_c$. Note that $T_v$ is
the claw tree $K_{1,c}$. It has no interior edges. These
definitions are illustrated in Figure 1.

\begin{figure}
\begin{center}  \includegraphics{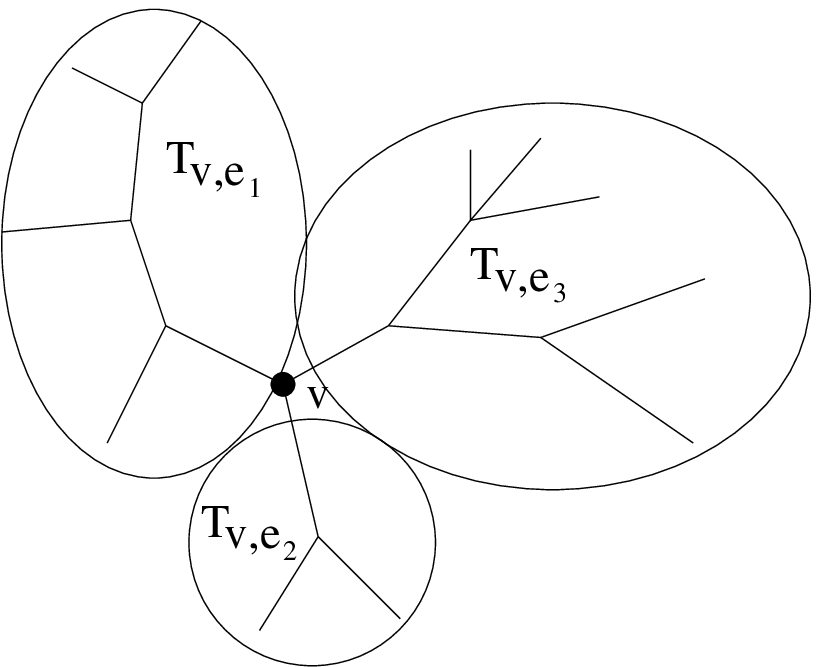}
$$\hbox{Figure 1: The subtrees around a vertex $v$ of the tree $T$} $$
\end{center} \end{figure}

\begin{lemma}\label{lem:localglobal}
Let $g $ be any binomial in the ideal  of the claw tree $T_v$,
written in tableau notation as
$$ g \quad = \quad \left[ \begin{array}{ccc}
l_1^1 &  \cdots & l_1^c \\
\vdots & \vdots & \vdots \\
l_d^1 & \cdots & l_d^c \end{array} \right] \, - \, \left[
\begin{array}{ccc}
m_1^1 &  \cdots & m_1^c \\
\vdots & \vdots & \vdots \\
m_d^1 & \cdots & m_d^c \end{array} \right] \quad \in \quad
I_{T_v,L}. $$
 For each row $i$ and column $j$, consider labelings $L^j_i \in
 {\rm im}(L^{T_{v,e_j}},l^j_i)$  and
 $M^j_i \in {\rm im}(L^{T_{v,e_j}},l^j_i)$ with the property
that the  multiset $\{L^j_i \}_{j=1}^d$ is equal to the multiset
$\{M^j_i \}_{j=1}^d$.
 Then the binomial
$$ g^*\quad = \quad  \left[ \begin{array}{ccc}
L_1^1 &  \cdots & L_1^c \\
\vdots & \vdots & \vdots \\
L_d^1 & \cdots & L_d^c \end{array} \right] - \left[
\begin{array}{ccc}
M_1^1 &  \cdots & M_1^c \\
\vdots & \vdots & \vdots \\
M_d^1 & \cdots & M_d^c \end{array} \right],$$
 belongs to toric ideal $I_{T,L}$ of the big tree $T$.
\end{lemma}

\begin{proof}
Since $L$ is friendly, each row in the tableaux is a consistent
labeling. Restricting to each subtree yields the same multiset of
labels. Hence the binomial $g^*$ is in $I_{T,L}$.
\end{proof}

\begin{defn}
Let $\mathcal{B}$ be any set of binomials in the (claw tree) ideal
$I_{T_v,L}$.  Denote by $\mathrm{Ext}(\mathcal{B} \rightarrow T)$
the set of all binomials $g^*$ gotten by applying the construction
in Lemma \ref{lem:localglobal} to the binomials $g \in
\mathcal{B}$.
\end{defn}

\begin{thm}[Local Structure of Invariants] \label{thm:local}
Let $T$ be a tree with a friendly labeling $L:G \rightarrow
\mathcal{L}$.  For each interior vertex $v$ of the tree $T$, let
$\,\mathcal{B}_v\,$ denote a binomial generating set for
$I_{T_v,L}$.  Then the following set of binomials generates the
ideal
 $I_{T,L}$ of all  phylogenetic invariants of $T$:
\begin{equation} \label{eqn:gen2}
\bigcup_v \mathrm{Ext}(\mathcal{B}_v \rightarrow T)\,\,\,\, \cup
\,\,\,\,\bigcup_e \mathrm{Quad}(e,T) .
\end{equation}
The first union is over the interior vertices of $T$. The second
union is over the interior edges of~$T$.
\end{thm}

\begin{proof} We proceed by
 induction on the number of interior vertices of $T$.  If there is
only one interior vertex then the statement is a tautology.
Suppose there are $m \geq 2$ interior vertices.  There exists an
interior vertex $v$ which is incident to only one other interior
vertex $u$.  Let $e$ be the edge connecting $v$ and $u$.  The tree
$T_{e,-}$ has $m- 1$ interior vertices and the tree $T_{e,+}$ has
only one interior vertex. By induction, the corresponding ideals
have generating sets that come in the form of (\ref{eqn:gen2}).
Applying Theorem \ref{thm:gromain} yields a generating set for
$I_{T,L}$ which is larger than the set of binomials listed in
(\ref{eqn:gen2}).  We claim that every binomial in the set
difference  (\ref{WeLikeThisSet})~$\backslash$~(\ref{eqn:gen2})
belongs to the ideal generated by (\ref{eqn:gen2}).  Indeed, each
such binomial differs from a binomial in (\ref{eqn:gen2}) by
swapping some of the labels in the columns corresponding to the
tree $T_{v,e}$. Such a swap can occur only when the edge label at
$e$ itself is the same for each row of the tableau involved in the
swap.  But such a swap (or sequence of such swaps) can be realized
by adding multiples of the quadratic binomials in ${\rm
Quad}(e,T)$.
\end{proof}

\section{The Toric Algebra of Group Multiplication}

The results of the previous section reduce the computation of our
toric ideals of  phylogenetic invariants to the local case,
namely, when the tree has only one interior node. Such a tree is a
claw tree $K_{1,n}$. The corresponding toric ideal $I_{G,n}$
depends only on two parameters: a finite (additive) abelian group
$G$ and a positive integer $n$. This construction furnishes a new
family of numerical invariants for any group $G$, and it may hence
be of
 independent interest to algebraists.

Throughout this section we assume that the labeling function $L$
is the identity map on a finite group $G$. Our object of interest
is the following monomial map between polynomial rings:
\begin{eqnarray*}
\mathbb{C}\bigl[q_{g_1,\ldots, g_n}\,:\, g_1 ,\ldots,g_n \in G
\bigr] \,\,\,\, \rightarrow & \mathbb{C}\bigl[\,
a^{(i)}_g \,:\, g \in G, \, i=1,\ldots,n\!+\! 1 \bigr] \\
q_{g_1, \ldots, g_n} \,\,\,\, \quad
 \mapsto  &
 a^{(1)}_{g_1} a^{(2)}_{g_2} \cdots a^{(n)}_{g_n}
  a^{(n+1)}_{g_1 + g_2 \cdots + g_n}
\end{eqnarray*}
Let $I_{G,n}$ denote the kernel of this ring homomorphism. This is
the ideal of phylogenetic invariants in the Fourier coordinates
for the claw tree $K_{1,n}$. Note that the definition of the toric
ideal $I_{G,n}$ makes sense for any group $G$, even if $G$ is not
abelian. It encodes the  group multiplication table. The following
example is the basic building block for the Kimura 3-parameter
model on a binary tree.

\begin{ex} \label{DNAclaw}
Let  $n=2$ and $\,G \,=\, \mathbb{Z}_2 \times  \mathbb{Z}_2$. We
identify the group elements with the nucleotides:
$$ A = (0,0), \,\, G = (0,1),  \,\, C = (1,0),  \,\, T = (1,1). $$
Then $\,I_{\mathbb{Z}_2 \times  \mathbb{Z}_2, 2}\,$ is an ideal in
$\, \mathbb{C}[
 q_{AA},  q_{AG},  q_{AC},  q_{AT},
 q_{GA},  q_{GG},  q_{GC},  q_{GT},
 q_{CA},  q_{CG},  q_{CC},  q_{CT},
 q_{TA}, $ $  q_{TG}, $ $  q_{TC},  q_{TT}]\,$.
It is the kernel of the monomial map $\, q_{g_1 g_2} \, \mapsto \,
x_{g_1} y_{g_2} z_{g_1+g_2}$. More specifically,
$$
q_{AA} \mapsto x_A y_A z_A ,\, \ldots \,,\, q_{AT} \mapsto x_A y_T
z_T \,,\, \ldots \,,\, q_{GC} \mapsto x_G y_C z_T \,,\,
\ldots\,,\, q_{TT} \mapsto x_T y_T z_A .  $$ The toric ideal
$\,I_{\mathbb{Z}_2 \times  \mathbb{Z}_2,2}\,$ is minimally
generated by the $16$ cubics
\begin{eqnarray*}
&  q_{AA} q_{CT} q_{TG} -  q_{AG} q_{CA} q_{TT},\,
 q_{AA} q_{GT} q_{TC} -  q_{AC} q_{GA} q_{TT},\,
 q_{AC} q_{CT} q_{TA} -  q_{AT} q_{CA} q_{TC},\, \\ &
 q_{AC} q_{GG} q_{TA} -  q_{AA} q_{GC} q_{TG},\,
 q_{AG} q_{CC} q_{TA} -  q_{AA} q_{CG} q_{TC} ,\,
 q_{AG} q_{GC} q_{CA} -  q_{AC} q_{GA} q_{CG},\, \\ &
 q_{AG} q_{GT} q_{CC} -  q_{AC} q_{GG} q_{CT},\,
 q_{AG} q_{GT} q_{TA} -  q_{AT} q_{GA} q_{TG},\,
 q_{AT} q_{CC} q_{TG} -  q_{AC} q_{CG} q_{TT},\, \\ &
 q_{AT} q_{GA} q_{CC} -  q_{AA} q_{GC} q_{CT},\,
 q_{AT} q_{GG} q_{CA} -  q_{AA} q_{GT} q_{CG},\,
 q_{AT} q_{GG} q_{TC} -  q_{AG} q_{GC} q_{TT},\, \\ &
 q_{GA} q_{CC} q_{TG} -  q_{GG} q_{CA} q_{TC},\,
 q_{GC} q_{CT} q_{TG} -  q_{GT} q_{CG} q_{TC},\,
 q_{GG} q_{CT} q_{TA} -  q_{GA} q_{CG} q_{TT},\, \\ &
 q_{GT} q_{CC} q_{TA} -  q_{GC} q_{CA} q_{TT}
\end{eqnarray*}
and the $18$ quartics
\begin{eqnarray*}  &
 q_{AA} q_{AT} q_{TG} q_{TC} -  q_{AG} q_{AC} q_{TA} q_{TT},\,
 q_{AA} q_{GG} q_{CT} q_{TC} -  q_{AG} q_{GA} q_{CC} q_{TT},\,\\ &
 q_{AA} q_{GT} q_{CC} q_{TG} -  q_{AC} q_{GG} q_{CA} q_{TT},\,
 q_{AA} q_{GT} q_{CT} q_{TA} -  q_{AT} q_{GA} q_{CA} q_{TT},\,\\ &
 q_{AC} q_{AT} q_{GA} q_{GG} -  q_{AA} q_{AG} q_{GC} q_{GT},\,
 q_{AC} q_{GA} q_{CC} q_{TA} -  q_{AA} q_{GC} q_{CA} q_{TC},\,\\  &
 q_{AC} q_{GA} q_{CT} q_{TG} -  q_{AG} q_{GT} q_{CA} q_{TC},\,
 q_{AC} q_{GT} q_{CG} q_{TA} -  q_{AT} q_{GC} q_{CA} q_{TG},\, \\ &
 q_{AG} q_{AT} q_{CA} q_{CC} -  q_{AA} q_{AC} q_{CG} q_{CT},\,
 q_{AG} q_{GC} q_{CC} q_{TG} -  q_{AC} q_{GG} q_{CG} q_{TC},\, \\ &
 q_{AG} q_{GC} q_{CT} q_{TA} -  q_{AT} q_{GA} q_{CG} q_{TC},\,
 q_{AG} q_{GG} q_{CA} q_{TA} -  q_{AA} q_{GA} q_{CG} q_{TG},\, \\ &
 q_{AT} q_{GG} q_{CC} q_{TA} -  q_{AA} q_{GC} q_{CG} q_{TT},\,
 q_{AT} q_{GG} q_{CT} q_{TG} -  q_{AG} q_{GT} q_{CG} q_{TT},\,  \\ &
 q_{AT} q_{GT} q_{CC} q_{TC} -  q_{AC} q_{GC} q_{CT} q_{TT},\,
 q_{CC} q_{CT} q_{TA} q_{TG} -  q_{CA} q_{CG} q_{TC} q_{TT},\, \\ &
 q_{GA} q_{GT} q_{CG} q_{CC} -  q_{GG} q_{GC} q_{CA} q_{CT},\,
 q_{GG} q_{GT} q_{TA} q_{TC} -  q_{GA} q_{GC} q_{TG} q_{TT}.
\end{eqnarray*}
Geometrically, these binomials define a $11$-dimensional toric
variety of degree $96$ in $\pp^{15}$. \qed
\end{ex}

Let $\phi(G,n)$ denote the largest degree of any minimal generator
of the toric ideal $I_{G,n}$. We computed these numbers for some
small groups $G$ and small values of $n$ using the toric algebra
software {\tt 4ti2} written by the Hemmeckes \cite{He}. The
results are displayed in the following table:

\smallskip

$$\begin{array}{|c|c||c|c|c|c|c||c|}
\hline G &\,\, n \,\, & 2 & 3 & 4 & 5 & 6 & \phi(G,n) \\
\hline\mathbb{Z}_2 & 2 & 0 & 0 & 0 & 0 & 0 & 0 \\
\hline\mathbb{Z}_2 & 3 & 3 & 0 & 0 & 0 & 0 & 2 \\
\hline\mathbb{Z}_2 & 4 & 30 & 0 & 0 & 0 & 0 & 2 \\
\hline\mathbb{Z}_2 & 5 & 195 & 0 & 0 & 0 & 0 & 2 \\
\hline\mathbb{Z}_2 & 6 & 1050 & 0 & 0 & 0 & 0 & 2 \\
\hline\mathbb{Z}_3 & 2 & 0 & 2 & 0 & 0 & 0 & 3 \\
\hline\mathbb{Z}_3 & 3 & 54  & 24  & 0 & 0 & 0 & 3 \\
\hline\mathbb{Z}_4 & 2 & 0 & 16 & 6  & 0 & 0 & 4 \\
\hline\mathbb{Z}_4 & 3 & 344  & 256  & 96  & 0 & 0 & 4 \\
\hline\mathbb{Z}_2 \times  \mathbb{Z}_2  & 2 & 0 & 16 & 18 & 0 & 0 & 4 \\
\hline\mathbb{Z}_2  \times  \mathbb{Z}_2 & 3 & 360  & 261  & 480 & 0 & 0 & 4 \\
\hline\mathbb{Z}_5 & 2 & 0 & 50  & 50  &  0 & 0 & 4 \\
\hline\mathbb{Z}_6 & 2 & 0 & 116 & 675 & 216 & 126 & 6 \\
\hline\mathbb{Z}_7 & 2 & 0 & 245  & 1764  & 1764 & 294 & 6 \\
\hline
\end{array}
$$

\medskip

The entry in the row labeled $(G,n)$ and  column labeled $i$ is
the number of minimal generators of $I_{G,n}$ having degree $i$.
For the two element group $\mathbb{Z}_2$ we can prove the
following general result:

\begin{thm} \label{z2quadrics}
The toric ideal $I_{\mathbb{Z}_2,n}$ is generated in degree two.
In symbols, $\phi(\mathbb{Z}_2,n) = 2$ for $n \geq 3$.
\end{thm}

\begin{proof}
Following the discussion in the previous section, the monomials in
the polynomial ring $\,\mathbb{C}\bigl[q_{g_1,\ldots, g_n}\,:\,g_i
\in \{0,1\} \bigr]\,$ are identified with tableaux. An $m \times
(n+1)$-tableaux $T$ with entries in $\{0,1\}$ represents a
monomial if and only if all row sums of $T$ are even. The ideal
$I_{\mathbb{Z}_2,n}$ is spanned by all binomials $T - T'$ where
$T$ and $T'$ are such tableaux which have the same column sums.

Consider any binomial $T-T'$  in the toric ideal
$I_{\mathbb{Z}_2,n}$. We can pick any two columns $i$ and $j$ and
switch each $0$ in these two columns to a $1$ and vice versa. The
resulting tableaux still have even row sums and their difference
is in $I_{\mathbb{Z}_2,n}$. We will use this symmetry in the next
paragraph.

Suppose that $I_{\mathbb{Z}_2,n}$ is not generated by quadrics.
Then the ideal contains a binomial $T - T'$ of degree $m \geq 3$
such that $T$ and $T'$ cannot be connected by moves involving only
two rows at a time. Such a move corresponds to adding a multiple
of a quadratic binomial.  We may suppose that $m$ is the smallest
degree of any such monomial. After permuting columns and applying
the symmetry described above, we may assume that
$$ T - T' \quad = \quad
 \left[ \begin{array}{cccccc}
0 & \cdots & 0 & 0 & \cdots & 0 \\
\vdots & \vdots & \vdots & \vdots & \vdots & \vdots  \\
 \end{array} \right] \,\,\, - \,\,
 \left[ \begin{array}{cccccc}
0 & \cdots & 0 & 1 & \cdots & 1 \\
\vdots & \vdots & \vdots & \vdots & \vdots & \vdots  \\
 \end{array} \right] . $$
We may further assume that the number $k$ of $1$'s in the first
row of $T'$ is less than or equal to the number of disagreements
between $T$ and $T'$ in any other row. The pair $(m,k)$ is thus
assumed to be lexicographically minimal among all such
counterexample binomials.

Consider the two rightmost columns. If there exists a pair $00$ in
these columns in tableau $T'$ then we can swap the pair $00$ with
the pair $11$ in the first row and get a counterexample with
smaller value of $m$. Likewise, if there exists a pair $11$ in
these columns in tableau $T$ then we can swap the pair $11$ with
the pair $00$ in the first row and get a counterexample with
smaller value of $m$. We thus conclude the sum of the two last
columns in $T'$ is at least $m+1$. Likewise, the sum of the two
last columns in $T$ is at most $m-1$. This is a contradiction to
the hypothesis that $T$ and $T'$ have the same column sums. This
completes the proof that  $I_{\mathbb{Z}_2,n}$ is generated by
quadrics.
\end{proof}

Theorem \ref{z2quadrics} and our computational results suggest the
following general conjecture.

\begin{conj} \label{atmostgrouporder}
For any finite abelian group $G$ and any positive integer $n$ we
have $\phi(G,n) \leq |G|$.
\end{conj}

If this conjecture holds then it is natural to define the
\emph{phylogenetic complexity} of a group $G$ as
$$ \phi(G) \,\, := \,\, {\rm max}_{n \geq 2} \, \, \phi(G,n)  .$$
The phylogenetic complexity $\phi(G)$ is an intrinsic invariant of
the group $G$. It makes perfect sense for arbitrary groups not
just abelian groups. However, if $G$ is not abelian then the
phylogenetic complexity can exceed the group order. Using the
software {\tt 4ti2}, we found that $\,\phi(S_3,2) \geq 8 \,$ for
the symmetric group on three letters.
 It would be interesting to
study the group-theoretic meaning of this invariant. For
applications in computational biology, however, it is the
four-element group of Example \ref{DNAclaw} which deserves the
most interest. We state this as a separate conjecture.

\begin{conj} \label{atmostfour}
The phylogenetic complexity $\phi(G)$ of the group $\,G \,=\,
\mathbb{Z}_2 \times  \mathbb{Z}_2\,$ is four.
\end{conj}

\section{Evolutionary Models for DNA Sequences}

Theorem \ref{cool} follows as a corollary from Theorem
\ref{thm:local} and the computational results for $n = 2$ in the
table of Section 5. In this section we make this explicit by
deriving the quadratic and cubic generators of the ideal of
phylogenetic invariants for the Jukes-Cantor models.  The
analogous derivation for the Kimura models will be sketched. Our
discussion is aimed at computational biologists who wish to work
with phylogenetic  invariants for evolution of DNA sequences.

\subsection{Specifying the root distribution}

The theory developed so far was based on the unrealistic
assumption that the structure of the root distribution is
constrained by the group structure associated to the transition
matrices.  In practice, the root distribution will be either the
uniform distribution or an arbitrary distribution.  In the first
case there are no parameters associated to the root and in the
second case there are $|G| - 1$ parameters associated to the root.
In either case, the setup differs slightly from that of Sections 3
and 4.  In this subsection we explain why all the results
including Theorem \ref{thm:gromain} still apply.

First we will suppose that, in our model, the root distribution
$\pi$ is the uniform distribution. Let $e_r$ be the corresponding
root edge.  By Lemma \ref{lem:fourier}, the Fourier transform of
the uniform distribution $\pi$ is the function that is equal to
one when evaluated at the identity and zero otherwise.  This means
that any Fourier coordinate $q_{g_1 \ldots g_m}$ with $g_1 +
\cdots+  g_m \neq 0$ is an invariant.

\begin{prop}[More Linear Invariants]
Fix $\pi$ to be the uniform distribution. Then the ideal $I_{T,L}$
consists of the previous invariants together with all linear
invariants $q_{g_1 \ldots g_m} $ with $g(e_r) \neq 0$.
\end{prop}

\begin{proof}
All of the theory we have developed for friendly labelings still
applies in this setting.  The only change is to restrict the set
of labels to the subset ${\rm im}(L^T, 0)$. This is the subset of
those labels $\lambda \in {\rm im}(L^T)$ which satisfy
 $\lambda(e_r) = 0$. Notice that  Theorem
\ref{thm:gromain} still applies since the ${\rm Ext}$ operator is
well-defined on sets of labels that are globally restricted on one
or more edge.
\end{proof}

Now consider the case where $\pi$ is allowed to be arbitrary in
the model under consideration.   In this case we are not
restricting the type of labels $\lambda$ which may appear, but  we
are in fact increasing the number and type of such labels.  The
labeling function $L$ is no longer the same on each edge of the
graph:  it is equal to the identity function on the edge
corresponding to the root distribution.  Such a mixed labeling
function need not be friendly everywhere.   However, it is still
friendly around any vertex that is not incident to the root edge.
More generally, if we consider any edge of the tree $e$ such that
the mixed labeling function $L$ is friendly on the tree $T_{e,-}$
and possibly unfriendly on the tree $T_{e,+}$, Theorem
\ref{thm:gromain} still applies since the binomials constructed by
the $\mathrm{Ext}$  operator are  valid binomials.  The crucial
result which guaranteed that these polynomials actually contained
unknowns which belonged to the ring was Lemma \ref{lem:labglue}.
Upon inspection of its proof, however, we see that this only
depended on $L$ being a labeling function that was friendly on
half of the tree:~$T_{e,-}$.

In summary, we can apply all of our constructive results in any of
the cases of biological interest,  regardless of whether or not
the root distribution is uniform or arbitrary.

\subsection{Jukes-Cantor binary model}

Let $T$ be a binary tree with $m$ leaves.  The Jukes-Cantor binary
model has transition matrices
$$ \begin{pmatrix}
b_v & a_v \\
a_v & b_v \end{pmatrix}.
$$
Here it is not necessary to require $a_v + b_v = 1$. We can regard
 $(a_v:b_v)$ as homogeneous coordinates.

We shall derive the invariants for this model in the Fourier
coordinates.  First assume that the root distribution is
 arbitrary. There are no linear invariants for this
model. Add an extra edge at the root to arrive at a new tree $T'$
with $m+1$ leaves.  According to Theorem \ref{thm:local} we need
to know the invariants from the tree $K_{1,3}$ at a vertex of $T'$
to determine a generating set for the ideal of all invariants
associated to this model.  However, a direct calculation shows
that there are no invariants associated to $K_{1,3}$ (this is the
first line of the table in Section 5).  So we only need to
consider the quadratic invariants associated to each edge of the
tree.  We now construct these explicitly.

The Fourier coordinates are $q_{g_1 \ldots g_{m+1}}$ where $g_i
\in \mathbb{Z}_2$ and $\sum g_i = 0$. These coordinates can be
identified with families of disjoint paths connecting leaves of
$T$. Consider any  interior edge $e$ of $T'$. We relabel the
leaves so that the split determined by the edge $e$ separates the
 leaves $1, 2, \ldots, j$ from the leaves $j+1, \ldots, m+1$.
 We construct two matrices $M_0$ and $M_1$ each having
$2^{j-1}$ rows and $2^{m-j}$ columns.  The rows of $M_i$ are
indexed by the sequences $(g_1, \ldots, g_j)$ such that $ g_1 +
\cdots + g_j  = i$ and the columns are indexed by the sequences
$(g_{j+1} , \ldots, g_{m+1})$ such that $g_{j+1} + \cdots +
g_{m+1} = i$.  The entry of $M_i$ in row $(g_1, \ldots, g_j)$ and
column $(g_{j+1} , \ldots, g_{m+1})$ is the indeterminate $q_{g_1
\ldots g_j g_{j+1} \ldots g_{m+1}}$.  The set
 $\,{\rm Quad}(e,T')\,$ is precisely the set of all $2 \times 2$
minors of the matrices $M_0$ and $M_1$.  Our generating set for
the ideal of invariants is the union the sets $\,{\rm
Quad}(e,T')\,$ as $e$ ranges over the interior edges. For the case
of the uniform root distribution, we add the invariants $q_{g_1
\ldots g_{m+1}} $ satisfying $g_{m+1} = 1$.

To obtain the ideal of invariants in the original probability
coordinates we apply the inverse Fourier transform.  In this
situation, this is the same as the \emph{Hadamard transform} which
appears frequently in the phylogenetics literature \cite{SS}. Each
Fourier coordinate gets replaced as follows:
$$q_{g_1 \ldots g_{m+1}} \quad = \quad \sum_{i_1, \ldots, i_m \in \mathbb{Z}_2}
(-1)^{g_1 \cdot i_1 + \cdots + g_m \cdot i_m}  \cdot p_{i_1 \ldots
i_m}$$

\begin{figure}[ht]
\begin{center}\includegraphics{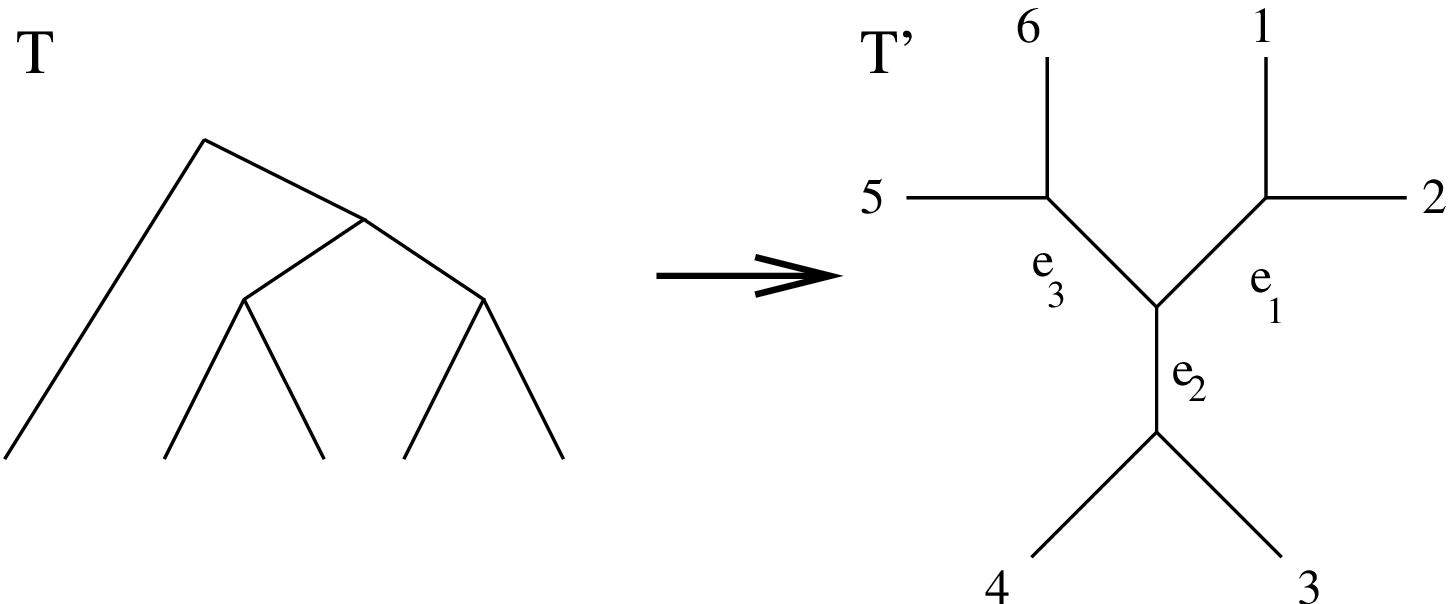}
$$ \hbox{Figure 2: Adding an edge
    at the root produces a snowflake} $$
\end{center}
\end{figure}

\begin{ex}[Snowflake]
Consider the tree $T$ on five leaves pictured in Figure 2. After
adding the extra edge at the root, we have the snowflake tree $T'$
with six leaves. Associated to each of the three interior edges
$e_1$, $e_2$, and $e_3$ there are 56 invariants which are the $2
\times 2$ minors of two $2 \times 8$ matrices.  For instance,
associated to the edge $e_1$ we get the two $2 \times 8$ matrices
$$ M_0 \quad = \quad  \begin{pmatrix}
q_{000000} & q_{000011} & q_{000101} & q_{001001} & q_{000110} &
q_{001010} & q_{001100} & q_{001111} \\
q_{110000} & q_{110011} & q_{110101} & q_{111001} & q_{110110} &
q_{111010} & q_{111100} & q_{111111} \\
\end{pmatrix} $$
$$ M_1 \quad = \quad \begin{pmatrix}
q_{010001} & q_{010010} & q_{010100} & q_{011000} & q_{010111} &
q_{011011} & q_{011101} & q_{011110} \\
q_{100001} & q_{100010} & q_{100100} & q_{101000} & q_{100111} &
q_{101011} & q_{101101} & q_{101110} \\ \end{pmatrix}.$$ A
probability distribution on five binary random variables comes
from the Jukes-Cantor binary model if and only if the $2 \times
2$-minors of all of  these six $2 \times 8$ matrices are zero.
\qed
\end{ex}

\subsection{Jukes-Cantor DNA model}

The Jukes-Cantor DNA model has transition matrices that look like
$$\begin{pmatrix}
b_v & a_v & a_v & a_v \\
a_v & b_v & a_v & a_v \\
a_v & a_v & b_v & a_v \\
a_v & a_v & a_v & b_v
\end{pmatrix}.  $$
Here it is not necessary to require $a_v + b_v = 1$. We can regard
 $(a_v:b_v)$ as homogeneous coordinates.
This is a group based model for $\, G = \mathbb{Z}_2 \times
\mathbb{Z}_2$ with the Jukes-Cantor labeling function $L : G
\rightarrow \{0,1\}$ defined  in Example \ref{jukeslabeling}.  As
was shown in \cite{SF} for binary trees and uniform root
distribution, the trees labeled by $L$ are precisely the
\emph{subforests} of $T$. There are  $F_{2m -1}$ subforests in a
binary tree with $m$ leaves, where $F_r$ is the $r$-th Fibonacci
number.  In total, for a tree with $m$ leaves, there are $3\cdot
4^{m-1}$ linear invariants of the form $q_{g_1 \ldots g_m}$ where
$g_1 + g_2 + \cdots + g_m \neq (0,0)$, and there are $4^{m-1} -
F_{2m -1}$ linear invariants of the form $\,q_{g_1 \ldots g_m} -
q_{h_1 \ldots h_m}\,$ where $L(g(e)) = L(h(e))$ for all $e$.

Now we will describe the higher degree invariants.  According to
Theorem \ref{thm:local} it suffices to understand the invariants
which arise for the (unrooted) claw tree $K_{1,3}$.  Modulo the
linear invariants, there are only five unknowns.  They correspond
to the five subforests of $K_{1,3}$ and they are
$$ \, q_{000}, \, q_{011}, \,   q_{101}, \,     q_{110},\, q_{111} .$$
 The phylogenetic ideal for this claw tree is generated by a single cubic
polynomial
$$ I_{K_{1,3},L} \quad = \quad \left< \,
q_{000} q_{111}^2 - q_{011} q_{101} q_{110} \, \right>. $$

 From this cubic we can deduce the ideal of invariants $I_{T,L}$
 provided $T$ is a binary tree.
  We express these invariants in the labeled coordinates
$q_\lambda$ where $\lambda$ is a sequence in $\{0,1\}^{|E(T)|}$
which is a consistent labeling of the tree $T$ according to the
Jukes-Cantor labeling function $L$.  That is, the $1$'s correspond
to the edges which appear in the corresponding subforest of $T$.

First, we will describe the quadratic binomials ${\rm Quad}(e,T)$
associated to each edge $e$. Form the matrix $M_0$ whose entries
are the
 unknowns $q_\lambda$ with $\lambda(e)  = 0$. The matrix
$M_0$ has $F_{2 m^- - 3}$ rows and $F_{2 m^+ - 3}$ columns where
$m^-$ is the number of leaves on $T_{e,-}$ and $m^+$ is the number
of leaves of $T_{e,+}$.  The rows (resp.~columns) of $M_0$ are
indexed by the subforests of $T_{e,-}$ (resp.~$T_{e,+}$) whose
labeling on $e$ is $0$.  Similarly, the matrix $M_1$ is a $F_{2m^-
- 2} \times F_{2m^+ - 2}$ matrix whose entries are the unknowns
$q_\lambda$ with $\lambda(e) = 1$.
  The set ${\rm Quad} (e,T)$ consists of all
the $2 \times 2$ minors of $M_0$ and $M_1$.

Now we will describe the cubic invariants associated to each
interior vertex  $v$ of the tree.  Let the three edges emanating
from $v$ be $e_1, e_2$ and $e_3$.  Recall that $T_{v,e_i}$ is the
subtree of $T$ which has $e_i$ as a leaf edge (see Figure 1).  The
set of consistent labels of $T_{v,e_i}$ that have label $l$ on
edge $e_i$ is denoted ${\rm im} ( L^{T_{v,e_i}},l)$.  This is just
the set of subforests on $T_{v,e_i}$ which have edge label $l$ on
the edge $e_i$.  Then we need to take all the cubic polynomials
derived from $\, q_{000} q_{111}^2 - q_{011} q_{101} q_{110}\,$ as
follows:
$$ q_{L^1_1, L^2_1, L^3_1} \cdot q_{L^1_2, L^2_2, L^3_2} \cdot
q_{L^1_3, L^2_3,L^3_3} \, \, - \,\,  q_{L^1_1, L^2_2, L^3_2} \cdot
 q_{L^1_2, L^2_1, L^3_3} \cdot q_{L^1_3, L^2_3,  L^3_1} , $$
where $\,L^i_1 \in {\rm im} ( L^{T_{v,e_i}},0)\,$,  $\,L^i_2,
L^i_3 \in {\rm im} ( L^{T_{v,e_i}},1)\,$, for all $i$. Now we will
illustrate how to apply these constructions on a small example.

\begin{figure}
\begin{center}
\includegraphics{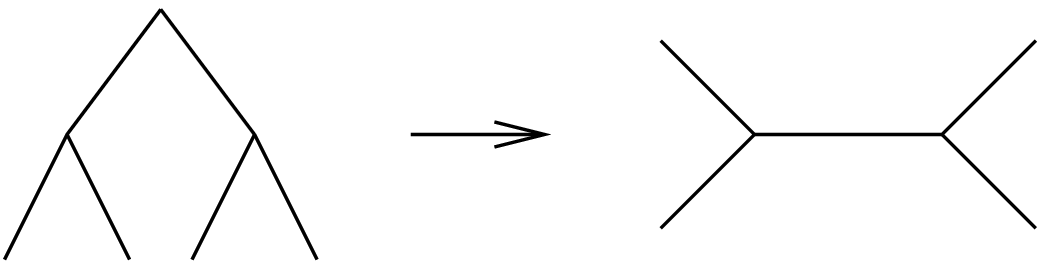}
$$ \hbox{Figure 3: The unrooted balanced binary
  tree when the root distribution is uniform} $$
\end{center}
\end{figure}

\begin{ex}[Balanced Binary Tree]
Let $T$ be the balanced binary tree with four leaves.  See Figure
3. We assume that the root distribution is uniform.  Modulo the
linear invariants there are $F_7 = 13$ indeterminates given by the
$13$ subforests of the binary tree with four leaves:
$$q_{00000}, q_{11000}, q_{00011}, q_{11011}, q_{10110}, q_{10101},
q_{01110}, q_{01101}, q_{11110}, q_{10111}, q_{01111}, q_{11101},
q_{11111}.$$ The first two indices in the label correspond to the
left-most leaves, the last two indices correspond to the
right-most leaves and the middle index is the interior edge. The
matrices $M_0$ and $M_1$ associated to the interior edge are
respectively the $F_3 \times F_3$ and $F_4 \times F_4$ matrices
$$M_0  \quad = \quad \begin{pmatrix}
q_{00000} & q_{00011} \\
q_{11000} & q_{11011} \end{pmatrix},
$$
\begin{equation}
\label{sankoff} M_1 \quad = \quad \begin{pmatrix}
q_{10110} & q_{10101} & q_{10111} \\
q_{01110} & q_{01101} & q_{01111} \\
q_{11110} & q_{11101} & q_{11111} \end{pmatrix}.
\end{equation}
The invariants ${\rm Quad} (e,T)$ are the $2 \times 2$ minors of
these matrices.  Among the cubic invariants associated to the left
interior vertex is the binomial $ \, q_{00011}q_{11110}q_{11101} -
q_{01110}q_{10101}q_{11011}$.
 \qed
\end{ex}

To construct the invariants when the root distribution is allowed
to be arbitrary amounts to changing the labeling function
associated to the root distribution to the identity labeling.
There are 11 labeled Fourier indeterminates modulo the linear
invariants.  A direct computation using the software {\tt 4ti2}
shows that the ideal of invariants is generated by $9$ quadrics
and $6$ cubics.

\subsection{Kimura 3-parameter model}

The Kimura 3-parameter model has transition matrices that looks
like
$$\begin{pmatrix}
d_v & a_v & b_v & c_v \\
a_v & d_v & c_v & b_v \\
b_v & c_v & d_v & a_v \\
c_v & b_v & a_v & d_v \end{pmatrix}$$ Here the labeling $L$ is the
identity function on $\,\mathbb{Z}_2 \times \mathbb{Z}_2 =
\{A,G,C,T\}$. The labeling being injective, there are no linear
invariants. We add a root edge to get a tree with one more leaf.
We first form the set of quadrics ${\rm Quad}(e,T)$ for each
interior edge $e$. They are the $2 \times 2$-minors of four
matrices $M_A, M_G,M_C,M_T$, one for each of the nucleotides which
may appear in the labeling on that edge. The next step is to
determine the set of local binomials $\ \mathrm{Ext}(\mathcal{B}_v
\rightarrow T)\,$ for any interior vertex $v$. The ingredients for
this are the $16$ cubics and the $18$ quartics displayed in
Example \ref{DNAclaw}. These $34$ binomials form a Gr\"obner basis
for the ideal of the claw tree $K_{1,3}$, and to each of them we
apply the extension procedure of Lemma \ref{lem:localglobal}.
Adding the resulting large collection of cubics and quartics to
the previous $2 \times 2$-minors gives generators for the ideal of
the Kimura 3-parameter model.

\subsection{Kimura 2-parameter model}

The Kimura 2-parameter model has transition matrices that look
like
$$\begin{pmatrix}
c_v & a_v & b_v & b_v \\
a_v & c_v & b_v & b_v \\
b_v & b_v & c_v & a_v \\
b_v & b_v & a_v & c_v \end{pmatrix}. $$ Here the group is also
$\,\mathbb{Z}_2 \times \mathbb{Z}_2 = \{A,G,C,T\}$, but the
labeling function is not injective. It is
$$ L:\mathbb{Z}_2 \times \mathbb{Z}_2
\rightarrow \{0,1,2\} \,\quad \hbox{with} \quad L(A) = 0, \,\,
L(G) = 1 \,\, \mbox{ and  } \, L(C) = L(T) = 2. $$ See Example
\ref{kimuralabeling}. Finding the set ${\rm im}(L^T)$ of
consistent labelings on a binary tree $T$ is a combinatorial
problem which we will not address here. (What is the analogue to
the Fibonacci numbers ?) Assuming this has been  accomplished and
the precise list of indeterminates $q_\lambda$ is known, then the
description of the set of quadrics $\,{\rm Quad}(e,T)$ associated
with an interior edge $e$ is just like before. They are the $2
\times 2$-minors of three matrices $\,M_0,M_1,M_2\,$ whose entries
are the unknowns $q_\lambda$.

In light of Theorem \ref{thm:local}, the remaining task is to
understand the ideal of invariants for the claw tree $K_{1,3}$.
Returning to the setup at the beginning of Section 3, this
 is an ideal in the 16 Fourier
coordinates $q_{g_1,g_2,g_3}$ where $g_1 + g_2 = g_3$.  There are
six linear invariants for this tree, which correspond to pairs of
triples $(g_1, g_2, g_3)$ and $(h_1, h_2, h_3)$ such that $L(g_i)
= L(h_i)$ for all $i$.   Modulo the linear invariants, the
polynomial ring has ten indeterminates $q_\lambda$. These are
$$q_{000}, q_{011}, q_{022},q_{101},q_{110},q_{122},
q_{202},q_{212},q_{220},q_{221}.$$
 The ideal of invariants for the claw tree $K_{1,3}$ modulo the linear
invariants has a Gr\"obner basis consisting of six cubics and
three quartics.  For example, the following two binomials appear
in it:
$$ q_{022}q_{101}q_{220} - q_{000}q_{122}q_{221} \quad \mbox{  and  }
 \quad  q_{022}^2q_{101}q_{110} -  q_{000}q_{011}q_{122}^2  . $$
Theorem \ref{thm:local} tells us how to construct the invariants
for any binary tree from these local data.

\section{Algebraically Independent Invariants Are Not Enough}

Each algebraic variety $X$ we have studied in  this paper lives in
an ambient space of $k^m$ dimensions, where $m$ is the number of
leaves of the given tree and $k$ is the number of states of each
random variable.  The coordinates of the ambient space are the
probabilities $p_{i_1 i_2 \cdots i_m}$, or their Fourier
transforms $q_{i_1 i_2 \cdots i_m}$. The \emph{dimension} of the
model $X$ is the number of free  model parameters, and the
\emph{codimension} of the model $X$ is
$$ {\rm codim}(X) \quad = \quad k^m \, - \, {\rm dim}(X). $$
This is the number of local equations needed to describe the
variety $X$ at a smooth point \cite{Ha}. However, in general, the
number of equations needed to describe $X$ at a singular point, or
the number of equations needed to define a variety $X$ globally,
can be much larger than the codimension.

Several research articles on  phylogenetic invariants give the
impression that to characterize a model $X$, it suffices to take
only ${\rm codim}(X)$ polynomial invariants, and some authors
raised   the question whether there is a complete list of
algebraically independent invariants. We wish to argue that, both
from the perspective of algebraic geometry and from the
perspective of computational biology, it is misleading and wrong
to ask for a set of only ${\rm codim}(X)$ polynomial invariants.

 Most models in
algebraic statistics, including the group-based evolutionary
models treated here, are \emph{not complete intersections}, i.e.,
these models require more polynomial equations than their
codimension. This holds even if one is only interested in strictly
positive probability distributions. In the opinion of the authors,
a given system of polynomial invariants for an evolutionary model
$X$ cannot be considered ``complete'' unless it actually generates
the
 prime ideal of $X$.

We illustrate this issue for the case when $X$ is the Jukes-Cantor
binary model (hence $k=2$) on the fully balanced binary tree with
$m=4$ leaves.  The parametric representation for this model was
given by (\ref{smallmap}). The variety $X$ has codimension $8$.
The homogeneous prime ideal of the model is given by the
 $2 \times 2$-minors of the four
$2 \times 4$-matrices in (\ref{twomatrices}). This ideal requires
$20$ minimal generators. Can we replace these $20$ quadrics by a
smaller subset? Don't eight suffice?

The answer is clearly ``no'' when $X$ is the complex variety
defined by requiring that the matrices (\ref{twomatrices}) have
rank one. However, more than eight equations are needed even if we
consider a small neighborhood of the centroid of the probability
simplex. This centroid is the uniform distribution on the leaf
colorations. In Fourier coordinates, this neighborhood is given by
setting $\, q_{0000} = 1\,$ and by assuming that the other $15$
coordinates $\, q_{ijkl} $ are real numbers of small absolute
value.

If we add the trivial invariant $\, q_{0000} - 1\,$ to our $20$
quadrics, then the resulting ideal in the polynomial ring in $15$
unknowns still has codimension $8$ but it is now minimally
generated by ten equations. The first five of these ten equations
express five of the unknowns in terms of the others:
$$ q_{1110} \, - \, q_{1100}  q_{0010}, \, q_{1111} \, - \, q_{1100}
q_{0011},\,  q_{1001} \, - \, q_{1000}  q_{0001}, \, q_{0111} \, -
\, q_{0011}  q_{0100}\,,    q_{1011} \, - \, q_{0011}  q_{1000}.
$$ What remains is an ideal of codimension $3$ which is minimally
generated by five quadrics. The five quadrics are the five $ 2
\times 2$-minors  not involving the upper left corner in the
following matrix:
$$ \begin{pmatrix}
            \bullet  & q_{0010} & q_{0001} \\
 q_{0100} & q_{0110} & q_{0101} \\
 q_{1000} & q_{1010} & q_{1001}
\end{pmatrix} $$
If we remove any of these five quadrics then the zero set of the
remaining four equations contains  points which are not in  the
model, even in a neighborhood of the uniform distribution. For
example, we get extraneous solutions by placing small positive
reals $\,\epsilon_{ijkl}\,$ in the matrices
$$ \begin{pmatrix}
            \, \bullet  & 0                   & 0  \\
            \,  0  & {\epsilon}_{0110} & {\epsilon}_{0101} \\
            \,  0 & {\epsilon}_{1010} & {\epsilon}_{1001}
\end{pmatrix}
\qquad \hbox{and} \quad
 \begin{pmatrix}
            \bullet  & {\epsilon}_{0010} & 0 \, \\
 {\epsilon}_{0100} & {\epsilon}_{0110} & 0 \, \\
 {\epsilon}_{1000} & {\epsilon}_{1010} & 0 \,
\end{pmatrix} $$
Notice that matrices with these entries are near the centroid of
the probability simplex and satisfy all but one of the five $2
\times 2$-minors of the matrix.  Thus we need all five quadrics to
define our variety, even set-theoretically, and even locally
around the uniform distribution. We regard the determinantal
formula (\ref{twomatrices}) as the best representation of the
ideal of phylogenetic invariants.

The failure to describe a phylogenetic model $X$ set-theoretically
becomes much more dramatic if we replace the ideal generators
derived in this paper with the \emph{canonical invariants}
introduced by Sz\'ekely, Steel and Erd\"os \cite{SSE}. The number
of canonical invariants is always equal to the codimension of $X$,
but, as we have argued, this means that they are far from having
the correct zero set. For the specific Jukes-Cantor binary model
with $m=4$ discussed above, there are eight canonical invariants.
From  \cite[Theorem 10]{SSE}, we see that they are the following
binomials of degree eight:
\begin{eqnarray*}
& q_{0000} q_{0010} q_{0100} q_{0110} q_{1001} q_{1011} q_{1101}
q_{1111} - q_{0001} q_{0011} q_{0101} q_{0111} q_{1000} q_{1010}
q_{1100} q_{1110} , \\  & q_{0000} q_{0010} q_{0101} q_{0111}
q_{1100} q_{1110} q_{1011} q_{1001} - q_{0001} q_{0011} q_{0100}
q_{0110} q_{1111} q_{1101} q_{1000} q_{1010} , \\  & q_{0000}
q_{0010} q_{0111} q_{0101} q_{1111} q_{1101} q_{1000} q_{1010} -
q_{0001} q_{0011} q_{0100} q_{0110} q_{1100} q_{1110} q_{1011}
q_{1001} , \\  & q_{0000} q_{0001} q_{0100} q_{0101} q_{1111}
q_{1110} q_{1011} q_{1010} - q_{0011} q_{0010} q_{0111} q_{0110}
q_{1100} q_{1101} q_{1000} q_{1001} , \\  & q_{0000} q_{0001}
q_{0111} q_{0110} q_{1111} q_{1110} q_{1000} q_{1001} - q_{0011}
q_{0010} q_{0100} q_{0101} q_{1100} q_{1111} q_{1111} q_{1010} ,
\\  & q_{0000} q_{0001} q_{0111} q_{0110} q_{1100} q_{1101}
q_{1001} q_{1010} - q_{0011} q_{0010} q_{0100} q_{0101} q_{1111}
q_{1110} q_{1000} q_{1001} , \\  & q_{0000} q_{0011} q_{0110}
q_{0101} q_{1110} q_{1101} q_{1000} q_{1011} - q_{0010} q_{0001}
q_{0100} q_{0111} q_{1100} q_{1111} q_{1010} q_{1001} , \\  &
q_{0000} q_{0011} q_{0100} q_{0111} q_{1110} q_{1101} q_{1010}
q_{1001} - q_{0010} q_{0001} q_{0110} q_{0101} q_{1100} q_{1111}
q_{1000} q_{1011}.
\end{eqnarray*}
The zero set of these equations has codimension three (!), and has
many irreducible components. The structure of the primary
decomposition of the ideal of canonical invariants  is very
complicated. For instance, among the irreducible components, there
are $48$ linear spaces of codimension $3$, e.g.
$$ q_{1111} \,\, = \,\, q_{1100} \,\, = \,\,
q_{1110} \,\, = \,\, 0 . $$ Among all the probability
distributions which satisfy the eight canonical invariants listed
above, the distributions which come from the Jukes-Cantor model
represent a subset that has measure zero (codimension $8$ inside
codimension $3$). For practical applications, this implies that an
empirical distribution which is near to the solution set of the
canonical equations cannot be trusted to come from the model.
Although the canonical invariants define the model locally almost
everywhere \emph{on the model distributions}, they do not define
the model globally in the entire probability simplex.

The canonical equations correspond to a lattice basis for the
toric ideal of phylogenetic invariants. It follows from general
theory in commutative algebra that the toric ideal can computed
from the canonical equations by the process of saturation (as
described in~\cite[Algorithm 12.3]{berndbook}), but this is a
non-trivial and time-consuming computation.  What we have
accomplished  in this paper is an explicit description of  a list
of phylogenetic invariants which minimally generates the toric
ideals of interest. This implies that globally (in the probability
simplex, in $\mathbb{R}^{k^m}$, or in $\cc^{k^m}$) the only points
which satisfy all the invariants come from the model.  However, in
all cases (with the exception of a few trivial ones), the number
of our polynomial invariants is considerably larger than the
codimension of the model, a feature which is unavoidable in
algebraic geometry.

There is another important motivation, coming directly from
computational biology, for our representation of the phylogenetic
invariants. Evolutionary models have to allow for the possibility
of \emph{heterogenous rates} as described in \cite{FS1, FS2}. For
instance, in the evolution of DNA sequences, one may wish to model
two different rates:  one for genes and one for non-genes. This
replaces our given parameterization (\ref{evolution1})
 by the superposition of two evolutionary models
 of the same kind:
$$
 p_{g_1 \ldots g_m} \quad = \quad
 \sum \, \pi_{g_r} \! \prod_{v \in \mathcal{V}(T)
  \setminus \{r\}} A^{(v)}_{g_{a(v)},g_v},
  \quad + \quad
 \sum \, \sigma_{g_r} \! \prod_{v \in \mathcal{V}(T)
  \setminus \{r\}} B^{(v)}_{g_{a(v)},g_v}.
  $$
  In statistics, this corresponds to introducing a hidden binary variable.
  In geometry,  we are passing to the
  \emph{secant variety} (see \cite[\S 7]{GSS}).
  Our determinantal
  presentation of the invariants ${\rm Quad}(e,T)$
  makes it easy to derive some invariants for models
  with heterogeneous rates. For instance, the
 cubic invariant discovered in \cite{FS2}
is nothing but the determinant of the $3 \times 3$-matrix in
(\ref{sankoff}).

\section{Conclusion}

This paper gives a solution to the longstanding problem of finding
all  phylogenetic invariants for the statistical models of
evolution which have a group structure. We found
 explicit Gr\"obner bases
for the ideals of the Jukes-Cantor and Kimura models for DNA
sequences. This was accomplished by developing a general machinery
for building invariants from the local features of a tree and
extending them to the entire tree.  There are, however, many
questions of a practical nature which remain.  The main issue is
how to use invariants to recover the phylogeny of a collection of
taxa.

First and foremost is the question of what statistical
significance should be attached to the numerical values that are
obtained by evaluating the phylogenetic invariants at sample data.
Intuitively, if the data come from the model associated to a
particular tree, the evaluation of an invariant polynomial should
be small.  How should this intuitive understanding be applied in
practice?  This is really a general open problem associated with
the polynomial functions that vanish on any statistical model.
The point of working with these polynomial invariants is that they
should eliminate the potentially difficult problem of
approximating solutions to the maximum likelihood equations.
However, most statistical tests (e.~g. $\chi^2$, $G^2$) depend on
comparing the empirical distribution to the maximum likelihood
estimates.  The fundamental open question we wish to pose to
statisticians is to develop statistical tests for deciding whether
or not the data fits a given model based solely on the evaluation
of the polynomials which vanish on the model distributions.

Even if the statistical issues in the previous paragraph can be
resolved, before we can start implementing a phylogeny recovery
method based on algebraic invariants, the help of computer
scientists is needed to address the following challenging
complexity question: \emph{How can we evaluate exponentially many
polynomials in exponentially many indeterminates for exponentially
many trees? }  The structural results about phylogenetic
invariants derived in this paper should help. For instance,  the
techniques of Section 4  will allow one to hunt for local features
of the tree (e.~g. 2- or 3-splits of the leaves) and assemble the
tree piece by piece.  Furthermore, our results show that \emph{all
quadratic phylogenetic invariants are rank conditions on matrices
associated to the splits of the tree}, so they can be interpreted
as conditional independence statements in the sense of graphical
models. These invariants are clearly well-suited for the
development of highly efficient algorithms.

Finally, now that we have explicit Gr\"obner bases for the
phylogenetic invariants of a group based model, there remains the
problem of determining how good invariant-based methods are at
recovering phylogenies in problems of interest to biologists.
Implementation and testing of invariant-based methods should be an
expanding area of future research, based on the work in this paper
and the results of Allman and Rhodes \cite{AR1,AR2} for the
general Markov model.

\section{Acknowledgements}
This work was supported by the National Science Foundation
(DMS-0200729). It was done while B.~Sturmfels held the Hewlett
Packard Visiting Research Professorship 2003-04 at MSRI~Berkeley.


\begin{thebibliography}{99}


\bibitem{AR1} E.~Allman and J.~Rhodes.  Phylogenetic  invariants for
  the general Markov model of sequence  mutation.
  \emph{Mathematical Biosciences}  {\bf 186} (2003) 133-144,


  \bibitem{AR2} E.~Allman and J.~Rhodes.  Quartets and parameter
  recovery for the general Markov model of sequence mutation,
  preprint, 2003.


\bibitem{CF} J.~A.~Cavender and J.~Felsenstein.
Invariants of phylogenies: a simple case with discrete states.
\emph{Journal of Classification} {\bf 4} (1987) 57-71.


\bibitem{ES} S.~Evans and T.~Speed.  Invariants of some probability
  models used in phylogenetic inference.  \emph{Annals of Statistics}
 {\bf  21}  (1993) 355-377.


\bibitem{EZ} S.~Evans and X.~Zhou.  Constructing and counting
  phylogenetic invariants.  \emph{Journal of Computational Biology}
  {\bf 5} (1998) 713-724.


\bibitem{Fel} J.~Felsenstein.  \emph{Inferring Phylogenies}.  Sinauer
  Associates, Inc., Sunderland, 2003.


 \bibitem{FS1} V.~Ferretti and D.~Sankoff.
Phylogenetic invariants for more general evolutionary models,
\emph{Journal of Theoretical Biology} {\bf 173} (1995) 147-162.


\bibitem{FS2} V.~Ferretti and D.~Sankoff.
A remarkable nonlinear invariant for evolution with heterogeneous
rates, \emph{Mathematical Biosciences} {\bf 134} (1996) 71-83.


\bibitem{GSS} L.~Garcia, M.~Stillman and B.~Sturmfels.
Algebraic geometry of Bayesian networks.
 \emph{Journal of Symbolic Computation}, to appear,
 {\tt math.AG/0301255}.


\bibitem{Ha} T.~Hagedorn.  Determining the number and structure of
  phylogenetic invariants.  \emph{Advances in Applied Mathemactics} {\bf 24} (2000) 1-21.


\bibitem{He} R.~Hemmecke and R.~Hemmecke. {\tt 4ti2} - Software for
  computing of Hilbert bases, Graver bases, toric Gr\"obner bases,
  and more.  Available at {\tt www.4ti2.de}, 2003.


\bibitem{HS} S.~Ho\c sten and S.~Sullivant. Gr\"obner bases and polyhedral
geometry of reducible and cyclic models.Ê \emph{Journal of
Combinatorial Theory: Series A} {\bf 100} (2002)  277-301.


\bibitem{La} J.~A.~Lake.  A rate-independent technique for
analysis of nucleaic acid sequences: evolutionary parsimony.
\emph{Molecular Biology and Evolution} {\bf 4} (1987) 167-191.


\bibitem{PS} L.~Pachter and B.~Sturmfels. The tropical geometry of
statistical models,  {\tt q-bio.QM/0311009}.


\bibitem{SB} D.~Sankoff and M.~Blanchette.
Phylogenetic invariants for genome rearrangements. \emph{Journal
of Computational Biology} {\bf 6} (1999) 431-445.


\bibitem{SS} C.~Semple and M.~Steel.  \emph{Phylogenetics}.  Oxford
  University Press, Oxford, 2003


\bibitem{SF} M.~Steel and Y.~Fu.  Classifying and counting linear
  phylogenetic invariants for the Jukes Cantor model.  \emph{Journal
  of Computational Biology} {\bf 2} (1995) 39-47.


\bibitem{SSEW} M.~Steel, L.~Sz\'{e}kely, P.~Erd\"{o}s, and P.~Waddell.
  A complete family of phylogenetic invariants for any number of taxa
  under Kimura's 3ST model.  \emph{NZ Journal of Botany} {\bf 13} (1993) 289-296.


\bibitem{berndbook}
B.~Sturmfels. \emph{Gr{\"o}bner Bases and Convex Polytopes},
American Mathematical Society, Providence, 1995.



\bibitem{SSE} L.~Sz\'{e}kely, M.~Steel, and P.~Erd\"{o}s.  Fourier
  calculus on evolutionary trees.  \emph{Advances in Applied Mathematics}
 {\bf 14} (1993) 200-216.


\end{thebibliography}
\end{document}